\documentclass[preprint,12pt]{elsarticle}

\usepackage{amssymb}
\usepackage{float}
\usepackage{amsmath}
\usepackage{bm} 
\usepackage[colorlinks=true, allcolors=blue]{hyperref} 
\usepackage{subcaption}
\usepackage{multirow}
\usepackage{cancel}
\usepackage[normalem]{ulem}
\usepackage[table]{xcolor}

\DeclareMathOperator{\sech}{sech}
\DeclareMathOperator{\sh}{sh}
\DeclareMathOperator{\ch}{ch}

\journal{Journal of Computational Physics}

\begin{document}

\begin{frontmatter}



\title{Unveiling the optimization process of Physics Informed Neural Networks: How accurate and competitive can PINNs be?}

\author[1]{Jorge F. Urbán\corref{cor}} 
\ead{jorgefrancisco.urban@ua.es}
\author[1,2]{Petros Stefanou}
\author[1]{José A. Pons}

\cortext[cor]{Corresponding author}

\affiliation[1]{organization={Departament de Física Aplicada, Universitat d'Alacant},
            addressline={Ap. Correus 99}, 
            city={Alacant},
            postcode={03830}, 
            state={Comunitat Valenciana},
            country={Spain}}
\affiliation[2]{organization={Departament d'Astronomia i Astrofísica, Universitat de València},
            addressline={Dr Moliner 50}, 
            city={València},
            postcode={46100}, 
            state={Comunitat Valenciana},
            country={Spain}}

\begin{abstract}

This study investigates the potential accuracy boundaries of physics-informed neural networks, contrasting their approach with previous  similar works and traditional numerical methods. We find that selecting improved optimization algorithms significantly enhances the accuracy of the results. Simple modifications to the loss function may also improve precision, offering an additional avenue for enhancement. Despite optimization algorithms having a greater impact on convergence than adjustments to the loss function, practical considerations often favor tweaking the latter due to ease of implementation. On a global scale, the integration of an enhanced optimizer and a marginally adjusted loss function enables a reduction in the loss function by several orders of magnitude across diverse physical problems. Consequently, our results obtained using compact networks (typically comprising 2 or 3 layers of 20-30 neurons) achieve accuracies comparable to finite difference schemes employing thousands of grid points. This study encourages the continued advancement of PINNs and associated optimization techniques for broader applications across various fields.

\end{abstract}


\begin{highlights}
\item We explore how the optimization process influences convergence and accuracy of Physics-Informed Neural Networks. 
\item We suggest adjustments to the BFGS algorithm and MSE loss to greatly enhance precision by orders of magnitude.
\item We explain our findings by analyzing the conditioning of the Hessian matrix and the spectrum of its eigenvalues.
\item We show that our scheme applies to various physical problems, offering greater accuracy and lower computational cost.

\end{highlights}


\end{frontmatter}



\section{Introduction}\label{sec:introduction}

Recent advances in physics-informed neural networks (PINNs) have positioned them as serious contenders in the domain of computational physics \cite{raissi2019physics,PINNS1}, disrupting the longstanding monopoly held by classical numerical methods in many and varied physical applications, such as fluid and solid mechanics \cite{sharma2023review, faroughi2024physics}, quantum optics \cite{ji2023}, black-hole spectroscopy \cite{2023PhRvD.107f4025L}, or radiative transfer \cite{2021JQSRT.27007705M}, among many others. This disruptive potential arises from their innate ability to integrate domain-specific physics principles with the powerful learning capabilities of neural networks. 

However, despite their significant potential, the nascent literature on PINNs is less than a decade old -which is minimal compared to numerical analysis, for example- and more often than desirable it suffers 
from a relative lack of rigorous mathematical analysis and specificity. The frequent reliance on trial-and-error methodologies, acquired from the machine learning literature, tends to obscure important insights into the origin of mathematical limitations, particularly concerning accuracy and efficiency.
Fortunately, as the field matures, an increasing number of studies are delving deeper into the mathematics underlying the components of PINNs. 

One of the most challenging aspects of neural networks is the inherently non-convex nature of their optimization problems. As a result, an increasing number of studies are delving deeper into their learning dynamics (see, for example, \cite{Smith2018SuperconvergenceVF, Lee_2020, Cohen2021GradientDO, 2022arXiv220714484C}). Focusing specifically on PINNs, \cite{WANGNTK} analyzes the convergence process using gradient descent methods by examining the eigenvalue spectra of the Neural Tangent Kernel (NTK).
The NTK has been discussed in many subsequent studies to provide a better understanding of the dynamics of the training process in many physical problems under different modifications of the standard PINN algorithm \cite{WANGFou, McClennyBragaNeto2023, Bai2023, Jha2024}. The learning dynamics of PINNs has recently been studied in \cite{GSNRPINNs} through the lens of the gradient signal-to-noise ratio (GSNR), which allowed the authors to characterize the optimization process of PINNs in different phases, as predicted by the Information Bottleneck theory \cite{IBtheory}. As a complementary path,
other works in the PINN literature have focused more on the development of new optimization techniques. In \cite{Ryck2023AnOP} they analyzed the behavior of gradient descent-based optimization methods, concluding that training with these algorithms is limited because of the ill-conditioning nature of the parameter space, and suggested various strategies to precondition this space. \cite{RLFLU24} introduced a novel second-order optimizer that significantly improves the solution returned by L-BFGS for some problems. Learnable optimization, also known in the field of deep learning as \emph{learning to learn} \cite{ChenLearnableOpt} has also been applied recently to PINNs in \cite{Bihlo2024}, improving the results obtained with standard machine learning optimizers.

Although PINNs exhibit remarkable adaptability to complex physical systems and have demonstrated interesting results across various domains, their efficacy is based on several factors that warrant closer scrutiny. One critical aspect is the architectural design of the neural network \cite{WangGradientPathologies, Wang2024PirateNetsPD, 2024arXiv240204390J, KANs}, which influences its ability to capture intricate physical phenomena accurately. Other ideas proposed to enhance the performance of PINNs are, for example, domain decompositions \cite{DomainDecomposition}, trainable activation functions \cite{JAGTAP2020109136}, loss function redefinitions \cite{WANGCausality}, or curriculum strategies \cite{krishnapriyan2021}, among others.

Despite their flexibility, training neural networks can be computationally expensive for many problems. The challenge also makes it difficult to determine which hyperparameters are crucial for improving results. Recent research has focused on this issue, particularly the weights controlling the residual and supervised terms, with new methods proposed to mitigate the negative effects of their magnitude disparity \cite{LIU2021112,WangTengPerdikarisGradientPathologies,WANG2024113112}. 
Some studies have also tried to give estimates of the errors as a function of different hyperparameters, 
such as the network size or the number of training points, for particular PDE problems \cite{Shin2020OnTC,Shin_2023,10.1093/imanum/drab093,10.1093/imanum/drac085,DeRyck2022,Biswas2022}. Scaling PINNs to large-scale or high-dimensional problems can be challenging. The neural network may require more layers or neurons, increasing the risk of overfitting or making the training process unstable. Tackling the so-called \emph{Curse of Dimensionality} is not a trivial task, and it has been recently studied in \cite{HU2024106369}.

Addressing these and other challenges needs a concerted effort from researchers to establish comprehensive benchmarks, standardized evaluation metrics, and theoretical frameworks that elucidate the fundamental principles governing the behavior of PINNs. Only through rigorous analysis and systematic experimentation can we unlock their full potential and turn them into reliable tools for solving complex physical problems across diverse fields, ranging from fluid dynamics and solid mechanics to quantum physics and beyond.

Although PINNs also have great potential in solving inverse problems (with applications in many and varied scenarios, such as fluid mechanics \cite{CMWYK2021,RYK2020,XWL2020,JMAK2022}, materials science \cite{ZDKS2022,SJBSK2022}, plasma physics \cite{MFHHZR2021}, among many others \cite{KYHWDP2020,2023MHPMM,ChenLuK2020}), which are crucial for many scientific and engineering applications, in this work, we focus on the use of PINNs for forward problems, specifically in solving partial differential equations for physics-related applications.
Nevertheless, we think that our findings could be also useful in other contexts, including inverse problems, which deserve a more thorough exploration.

This paper aims to improve our understanding of the fundamental aspects that define PINNs performance. At its core, training reduces to an optimization problem, prompting us to revisit the basics of optimization theory. We explore some intricacies of PINN optimization, seeking to identify the bottleneck that hinders their precision in various physical applications. Our focus extends beyond mere adjustments in network size, architecture, activation functions, or other hyperparameters. Instead, we argue that perhaps the most important ingredients are the fundamental principles that govern the optimization process. Through the exploration of different techniques, we bracket the boundaries of precision achievable with PINNs across diverse physical scenarios, focusing on the pivotal role played by the choice of the optimizer in determining the accuracy and efficiency of PINN solutions.

Through our investigation, we demonstrate that seemingly minor modifications in the optimization process can yield substantial enhancements in accuracy, often spanning orders of magnitude. By fine-tuning the optimizer selection, we uncover improvements that result in refined solutions with high precision. Moreover, this allows us to reduce the size of the network in tandem with the choice of optimizer, resulting in significant reductions in computational overhead. This focus-in-optimization approach not only enhances accuracy but also saves computational resources, paving the way for faster and more efficient simulations with PINNs, and enhancing their scalability and applicability across diverse domains of physics and engineering.

The paper is structured as follows: Section \ref{sec:summary} provides a brief overview of our PINN framework and discusses key issues related to commonly used optimizers. We dive into the details behind the optimization process in PINNs and we present our proposed contributions to improve it in Section \ref{sec:optimization_method}. In Section \ref{sec:ns_magnetosphere}, we thoroughly examine a relatively simple case to demonstrate the significant impact of selecting the appropriate optimizer and how such a choice can minimize network size while achieving excellent results, surpassing previous studies with similar problems. Additionally, in Section \ref{sec:other_problems}, we present a comprehensive set of physical problems spanning various fields, illustrating that the insights from the preceding section can be extended to a large variety of problems. Finally, Section \ref{sec:conclusions} summarizes our findings and outlines potential paths for future improvements. All code accompanying this manuscript is available on GitHub at \url{https://github.com/jorgeurban/self_scaled_algorithms_pinns}.

\section{Summary of the PINNs approach.}\label{sec:summary}

Given a set of coordinates $x_\alpha = (x_1, x_2, x_3, ...)$ in some domain $D$, a general partial differential equation (PDE) that describes the state $u$ of a physical system can be written in the form
\begin{align}
    \mathcal{L} u(x_\alpha) &= G(x_\alpha, u(x_\alpha)), \label{eq:general_pde}
\end{align}
where $\mathcal{L}$ is a non-linear differential operator and $G$ is a source term.
The PINN approach, as introduced by \cite{Lagaris} and \cite{raissi2019physics}, solves the problem by finding a neural network surrogate $u(x_\alpha; \bm{\Theta})$ that approximates the true solution of the problem. 
The set of parameters $\bm{\Theta}$ (i.e. the weights and biases of the neural network) is adjusted iteratively through an optimization process, which tries to minimize a loss function $\mathcal{J}$ that reflects a global measure of how well is equation \eqref{eq:general_pde} satisfied.
Typically, the loss function is defined as the mean squared error (MSE) of the residuals for a large number of points $N$
\begin{equation}\label{eq:loss_function}
    \mathcal{J} = \frac{1}{N}\sum_{i=1}^N | \mathcal{L} u(x_{\alpha i};\bm{\Theta}) - G(x_{\alpha i}, u(x_{\alpha i};\bm{\Theta}))|^2 ~.
\end{equation}

To completely describe the physical system, we must impose boundary conditions on a boundary $\partial D$.
When Dirichlet and periodic boundary conditions are involved, we impose them through the so-called {\it hard-enforcement} \cite{Lagaris, DN21, Luetal, Sethi2023, Sukumar2022}. This means that we redefine the solution so that they are satisfied by construction independently of the PINN's output.
In the case of Dirichlet boundary conditions, this can be achieved 
by redefining the solution in the following way:
\begin{equation}\label{eq:hard_enforcement_dirichlet}
    u(x_\alpha; \bm{\Theta}) = f_b(x_\alpha) + h_b(x_\alpha) 
    \mathcal{N}(x_\alpha,\bm{\Theta}),
\end{equation}
where $f_b$ is a suitable smooth function that satisfies the Dirichlet boundary conditions when $x_\alpha \in \partial D$, $h_b$ is a suitable smooth function that vanishes when $x_\alpha \in \partial D$, and $\mathcal{N}$ is the output of the PINN.

Periodic boundary conditions with periodicity $L$ can also be hard-enforced, as amply discussed in \cite{DongNi}. For example,
let us designate $x_\beta$ and $x_\gamma$ as subsets of $\partial D$ where we apply Dirichlet and periodic boundary conditions, respectively. Then, if we redefine the solution as follows:
\begin{align}
    u(x_\alpha; \bm{\Theta}) &= f_b(x_\beta) 
    + h_b(x_\beta) ~ \mathcal{N}(x_\beta, \cos\left({k x_\gamma}\right), \sin\left({k x_\gamma}\right) ; \bm{\Theta}), \label{eq:hard_enforcement_periodic} 
\end{align}
where $k = 2 \pi/L$,
one can indeed check that $u$ has the desired behavior at all the borders. This parametrization enforces periodicity for $u$ and all its derivatives with respect to $x_\gamma$, provided that $f_b$ and all its derivatives are also periodic. However, this does not necessarily work for all cases (for example, if we want a periodic function but not its derivatives). For some problems, the full Fourier expansion is formally needed to compute the solution (see e.g. \cite{ANAGNOSTOPOULOS2024116805,Wang2024PirateNetsPD, Hao2024}). Another possible approach is to consider a different set of functions, such as Hermite-based interpolation polynomials \cite{DongNi}, chosen to be periodic up to the desired order. These polynomials contain additional trainable parameters to enforce the appropriate relation between the function and the derivatives at the boundaries.

For Neumann or Robin boundary conditions, hard-enforcement is also possible but not straightforward to implement and can result in rather cumbersome and complicated expressions.
In these cases, one can always use the soft-enforcement, i.e. add terms in the loss function \eqref{eq:loss_function} that take into account the residuals of the boundary condition on a sample of points at the boundary.
The soft-enforcement approach is highly flexible and can be applied to all the previously discussed cases. However, its main drawback is that its performance depends on selecting additional hyperparameters, such as the number of boundary points and the relative weights assigned to these terms, which influence the focus on boundary conditions during the optimization process.

The loss function $\mathcal{J}$ is a multidimensional scalar function of the parameters $\bm{\Theta}$. 
Its minimization requires a robust optimization algorithm that updates the parameters after each training iteration. 
Two very popular choices in the literature of PINNs are the Adam \cite{ADAM} and BFGS \cite{BroydenBFGS,FletcherBFGS,Goldfarb1970AFO,ShannoBFGS} optimizers. 
The Adam optimizer has consistently been a fundamental component in the training of various machine learning applications and PINNs. However, it has recently become clear that quasi-Newton methods such as BFGS or its low-memory variant L-BFGS \cite{LiuNocedal} can achieve more accurate results in significantly fewer iterations than Adam, but they are more prone to be trapped at saddle points. The state-of-the-art training schemes involve a combination of these two optimizers, using Adam for the initial iterations to handle better the possible presence of saddle points, and then using BFGS / L-BFGS to accelerate convergence.

\section{Optimization method}\label{sec:optimization_method}
\subsection{Brief review of optimization procedures}
At this point, it is worth reviewing the basic concepts of optimization theory.
Both of the aforementioned optimizers can be encompassed in the general family of Line Search methods \citep{NoceWrig06}, where the iterative procedure consists of updating the parameters as follows:
\begin{align} 
    \bm{\Theta}_{k+1} &= \bm{\Theta}_k + \alpha_k \bm{p}_k, \label{eq:next_iterate}
\end{align}
where $\bm{p}_k$ is the direction of the correction step, which depends on the gradient of the loss function and some symmetric matrix $H_k$
\begin{align}
    \bm{p}_k &= -H_k \nabla \mathcal{J}(\bm{\Theta}_k), \label{eq:direction}    
\end{align}
and $\alpha_k$ is the step size, which varies depending on the particular method. 
The parameter $\alpha_k$ needs to be appropriately chosen to ensure the accuracy of the local gradient estimation and facilitate the loss function reduction, but without impeding convergence by being excessively small.

The simplest case is to consider the gradient decent algorithm, where $H_k = I$ and $\alpha_k$ is equal to a small positive constant, which has linear convergence. The Adam optimizer can be recovered by using $H_k = I$ and a formula for calculating a specific $\alpha_k$ for each parameter. It can be seen as a more sophisticated variant of the gradient descent algorithm, which has better convergence properties but is still linear.
Newton's method can be recovered by considering $H_k$ to be the exact inverse of the Hessian matrix of $\mathcal{J}$.
It is a second-order method that can converge in very few iterations but with a large increase in the computational cost of each iteration: it requires the explicit calculation of second derivatives of the loss function to calculate the Hessian matrix, and then also its inversion, which can be prohibitively costly in large-scale optimization problems. Moreover, in non-convex problems the Newton's method could give away from the minimum non-descent directions, ultimately leading to line search failure.
The so-called Quasi-Newton methods lie in between. They use some approximation of the inverse Hessian that requires only the first derivatives of the loss function and involves only matrix-vector multiplications, resulting in superlinear convergence (not yet quadratic), but are much faster than Newton's method per iteration. The step size $\alpha_k$ is usually chosen with inexact line search procedures that preserve the positive-definiteness of $H_k$ by imposing certain restrictions on $\alpha_k$ (for example, the Wolfe conditions \cite{Wolfe}).

A general class of quasi-Newton iteration algorithms can be casted under the \textit{self-scaled Broyden} formula.
If we define the auxiliary variables 
\begin{align}
    \bm{s}_k &= \bm{\Theta}_{k+1} - \bm{\Theta}_k, \label{eq:sk} \\
    \bm{y}_k &= \nabla \mathcal{J}(\bm{\Theta}_{k + 1}) - \nabla \mathcal{J}(\bm{\Theta}_k), \\
    \bm{v}_k &= \sqrt{\bm{y}_k \cdot H_k \bm{y}_k}\left[ \frac{\bm{s}_k}{\bm{y}_k \cdot \bm{s}_k} - \frac{H_k \bm{y}_k}{\bm{y}_k \cdot H_k \bm{y}_k}\right],
\end{align}
the next approximation of the inverse Hessian matrix at each iteration can be calculated by (see \cite{Al-Baali1993} and \cite{AlBaaliKhalfan}) 
\begin{equation}\label{eq:self_scaled_broyden}
    H_{k+1} = \frac{1}{\tau_k}\left[ H_k - \frac{H_k \bm{y}_k \otimes H_k \bm{y}_k}{\bm{y}_k \cdot H_k \bm{y}_k} + \phi_k \bm{v}_k \otimes \bm{v}_k \right] + \frac{\bm{s}_k \otimes \bm{s}_k}{\bm{y}_k \cdot \bm{s}_k}, 
\end{equation}
where $\otimes$ denotes the tensor product of two vectors and $\tau_k, \phi_k$ are respectively the scaling and the updating parameters, which in general change between iterations. For $\tau_k = 1$ and $\phi_k = 1$, we recover the standard BFGS algorithm.

\subsection{Modifications of optimization algorithm}
In this work, we introduce two methods, which we label as self-scaled BFGS (SSBFGS) and self-scaled Broyden (SSBroyden) method respectively.
These methods are well-established in optimization theory and have demonstrated certain advantages \cite{AlBaaliKhalfan, Al-Baali1}.
In fact, they could be considered as modifications to the BFGS formula rather than new, standalone optimizers.
The term ``self-scaled" means that $\tau_k \neq 1$ and corresponds to the usual BFGS formula, but with a scaling factor multiplying the approximation $H_k$ of the inverse Hessian, while ``Broyden" method assumes $\phi_k \neq 1$.

For SSBFGS we use the choices suggested in \cite{Al-Baali1}:
\begin{align}\label{eq:ssbfgs}
    \tau_k^{(1)} &=\min \left \lbrace 1, \frac{\bm{y}_k \cdot \bm{s}_k}{\bm{s}_k \cdot H_k^{-1} \bm{s}_k} \right \rbrace \\
    \phi_k &= 1.
\end{align}
In \ref{app:tau_k} we elaborate on the details of this optimizer and show how $\tau_k^{(1)}$ can be efficiently calculated without the need to invert the matrix $H_k$.

For SSBroyden we use the choices suggested in \cite{AlBaaliKhalfan}:
\begin{align}
    \tau_k^{(2)} &= \begin{cases}
                        \tau_k^{(1)} \min \left(\sigma_k^{-1/(n-1)}, \frac{1}{\theta_k} \right) 
                        & \quad \mathrm{ if } ~ \theta_k > 0 \\
                        \min \left( \tau_k^{(1)} \sigma_k^{-1/(n-1)}, \sigma_k \right) 
                        & \quad \mathrm{ if } ~ \theta_k \leq 0, 
                    \end{cases} \label{eq:tau2} \\ 
    \phi_k^{(1)} &= \frac{1 - \theta_k}{1 + a_k \theta_k}, \label{eq:phi1}
\end{align}
where $n = \mathrm{size}({\bm\Theta}_k)$ is the total number of the trainable parameters and  $\sigma_k, \theta_k, a_k$ are intermediate auxiliary variables,
defined through the following relations:
\begin{align} 
    b_k &= \frac{\bm{s}_k \cdot H_k^{-1} \bm{s}_k}{\bm{y}_k \cdot \bm{s}_k} = -\alpha_k \frac{ \bm{s}_k \cdot \nabla \mathcal{J}\left(\bm{\Theta}_k\right)}{\bm{y}_k \cdot \bm{s}_k}, \\
    h_k &= \frac{\bm{y}_k \cdot H_k \bm{y}_k}{\bm{y}_k \cdot \bm{s}_k}, \\
    a_k &= h_k b_k -1 \\
    c_k &= \sqrt{\frac{a_k}{a_k + 1}}, \\
    \rho_k^{-} &= \min \left(1, h_k \left(1 - c_k \right) \right), \\
    \theta_k^{-} &= \frac{\rho_k^{-}-1}{a_k}, \\
    \theta_k^{+} &= \frac{1}{\rho_k^{-}}, \\
    \theta_k &= \max \left(\theta_k^{-}, \min 
    \left(\theta_k^{+}, {\frac{1 - b_k}{b_k}}\right) \right) \\
    \sigma_k &= 1 + a_k\theta_k.
\end{align}
We should stress here that $\tau_k$ should respect certain restrictions in order to ensure global and super-linear convergence of the updating algorithm. In particular, for any $\theta_k$ that satisfies the inequality $(1-\nu_1) (-\frac{1}{a_k}) \leq \theta_k \leq \nu_2 < \infty$, there exists a $\tau_k$ for which the inequalities $(1-\nu_1) (-\frac{1}{a_k}) \leq \tau_k \theta_k \leq 1 - \nu_3, \nu_4 \leq \tau_k \leq 1$ hold, with $\nu_i > 0$ constants associated with the machine accuracy \cite{Al-Baali1998a}.
Furthermore, the choices presented here are not the only possible ones.
We extensively experimented with different options and determined that these choices consistently led to improved convergence and more precise solutions across the diverse range of problems we explored. For a comprehensive review with numerous references regarding diverse quasi-Newton methods within the self-scaled Broyden family, see \cite{BroydenMethodsReview}.

\subsection{Modifications of the loss function}\label{sec:loss_function}

Apart from improvements in the optimization algorithm, we also investigate the effect of using a slightly modified version of the usual MSE loss function \eqref{eq:loss_function}.
In particular, we explore the consequences of evaluating the loss function through a user-defined monotonically increasing function $g$:
\begin{equation}\label{eq:g_loss_function}
    \mathcal{J}_g = g \left( \mathcal{J} \right).
\end{equation}
The Hessian matrices $\mathrm{hess}(\mathcal{J})$ and $\mathrm{hess}(\mathcal{J}_g)$, associated respectively with the functions $\mathcal{J}$ and $\mathcal{J}_g$, are related by
\begin{equation}\label{eq:g_hess}
    \mathrm{hess}(\mathcal{J}_g) = g'(\mathcal{J}) \mathrm{hess}(\mathcal{J}) + g''(\mathcal{J}) \nabla \mathcal{J} \otimes \nabla \mathcal{J}.
\end{equation}
Near the minimum, the second term can be neglected, because $\nabla \mathcal{J} \simeq 0$, and both Hessian matrices are proportionally related to each other with ratio $g'(\mathcal{J})$.
Two obvious choices for $\mathcal{J}_g$ are 
\begin{align}
    \mathcal{J}_{1/2} &\equiv \sqrt{\mathcal{J}}, \label{eq:sqrt_MSE} \\
    \mathcal{J}_{\log} &\equiv \log{\mathcal{J}} \label{eq:logMSE}.
\end{align} 
Since both the square root and the natural logarithm exhibit derivatives exceeding 1 when $\mathcal{J} \ll 1$, they have the potential to accelerate convergence.

\section{Case study: neutron star magnetospheres}\label{sec:ns_magnetosphere}

We now focus on the problem of force-free neutron star magnetospheres in the non-rotating axisymmetric regime as a baseline case.
This problem was examined in detail in \cite{USDP23}, illustrating the potential of the PINN approach in this particular astrophysical scenario.
Here, we revisit this study to highlight the significant influence that the selection of the optimization algorithm and/or loss function can have on performance. While a detailed exposition of the theoretical background can be found in the aforementioned paper, we provide a brief overview of the core concepts and equations in \ref{app:gs_derivation} for completeness.

It is convenient to use compactified spherical coordinates $x_\alpha = (q, \mu, \phi)$, where $q=1/r$ and $\mu = \cos{\theta}$ and introduce dimensionless units $R$ (radius of the star) and $B_0$ (surface magnetic field at the equator of the dipolar component). In axisymmetry, the magnetic field $\bm{B}$ can be described in terms of two poloidal and toroidal stream functions $\mathcal{P}$ and $\mathcal{T}$. The problem then reduces to the so-called Grad-Shafranov equation
\begin{equation}\label{eq:grad_shafranov}
     \triangle_{\mathrm{GS}} \mathcal{P} + \mathcal{T} \frac{d\mathcal{T}}{d\mathcal{P}} = 0~,
\end{equation}
where we have defined the second order differential operator
\begin{align}
    \triangle_{\mathrm{GS}} &= \nabla \cdot \left(\frac{q^2}{1 - \mu^2} \nabla \right) \nonumber\\
     &= q^2 \left( q^2 \frac{\partial^2}{\partial q^2} + 2q \frac{\partial}{\partial q}\right) + q^2 \left(1-\mu^2 \right) \frac{\partial^2}{\partial \mu^2}.
\end{align}

In the general form for the loss function given by equation \eqref{eq:loss_function}, we can identify $\mathcal{L} = \triangle_{\mathrm{GS}}$, $G = -\mathcal{T} \frac{d\mathcal{T}}{d\mathcal{P}}$.  

We hard-enforce boundary conditions using equation \eqref{eq:hard_enforcement_dirichlet} with
\begin{align}
    f_b (q, \mu) &= q \left(1-\mu^2\right) \sum_{l=1}^{l_{\mathrm{max}}} b_l P_l'(\mu)\label{eq:surface_bc} \\
    h_b (q, \mu) &= q(q-1)(1-\mu^2),
\end{align}
where $P_l$ are the Legendre polynomials, $b_l$ are appropriate coefficients describing the solution at the surface of the star and prime denotes differentiation with respect to $\mu$.
This reformulation guarantees that $\mathcal{P}$ equals zero at the axis $(\mu = \pm 1)$, vanishes at infinity $(q = 0)$, and precisely fulfills the Dirichlet boundary condition at the surface $(q = 1)$.

\subsection{Current-free Grad-Shafranov equation (CFGS)}\label{sec:current_free_gs}

We begin by considering a current-free magnetosphere with $\mathcal{T}(\mathcal{P}) = 0$. Then, equation \eqref{eq:grad_shafranov} has an analytical solution given by
\begin{equation}\label{eq:analytical_GS}
    \mathcal{P}_\mathrm{an} (q,\mu) = (1-\mu^2)\sum_{l=1}^{l_\mathrm{max}}{q^l b_l P_l^{'}(\mu)},
\end{equation}
which is completely determined by the surface boundary conditions (providing the $b_l$ coefficients).
This is a relatively simple problem to solve but, nevertheless, valuable conclusions can be drawn for the optimization procedure by analyzing it and comparing it with the analytical solution.
The complexity of the problem depends on the number of multipoles considered in \eqref{eq:analytical_GS}, i.e. on the number of non-zero coefficients $b_l$.
Simpler solutions will achieve the same accuracy with fewer trainable parameters compared to more complex ones. To illustrate this, we focus on a dipole-quadrupole solution ($b_1, b_2 \neq 0, b_{l>2} = 0$).
Subsequently, we will extend our findings to encompass a broader range of force-free solutions and various problem types.

\begin{table}[]
    \centering
    \small
        \caption{Architecture and training hyperparameters for the two cases considered in this section. In both cases, we use a $\tanh$ activation function for the hidden layers. The \textit{Neurons} column refers to neurons per hidden layer. The \textit{Adam it.} column refers to the number of iterations where the Adam optimizer is used before switching to a quasi-Newton method. The \textit{Batch size} column refers to the number of points sampling the domain for each training set. The training set changes every 500 iterations in order to sample as many points as possible.}
    \begin{tabular}{c c c c c c c}
        \hline 
        PDE & Layers & Neurons & Iterations & Adam it. & Batch size & Domain 
        \\   & & & (x1000) & (x1000)& (x1000) &
        \\ [0.5ex]
        \hline 
        CFGS & 1 & 30 & 5 & 2 & 1 & $[0,1] \times [-1,1]$  \\ [0.5ex]
        NLGS & 2 & 30 & 20 & 10 & 8 & $[0,1] \times [-1,1]$ \\ [0.5ex]
        \hline
    \end{tabular}
    \label{tab:training_params}
\end{table}

\subsubsection{Impact of the optimization algorithm}\label{sec:optimization_algorithms}

The architecture and training hyperparameters that we use are outlined in table \ref{tab:training_params}.
We address the problem utilizing four different optimization methods: Adam, BFGS, SSBFGS and SSBroyden. 

We refresh the randomly sampled training points every 500 iterations. We found this simple approach effective, providing a good balance between loss reduction and avoiding overfitting. While more advanced sampling techniques could be used to further improve results (see \cite{Wu2023} and references for details), we stick to this simpler method to isolate and focus on the impact of our proposed methodology: without other refinements, variations in optimizers will lead to improvements in precision of several orders of magnitude.

\begin{figure}
    \centering
    \includegraphics[width=.999\textwidth]{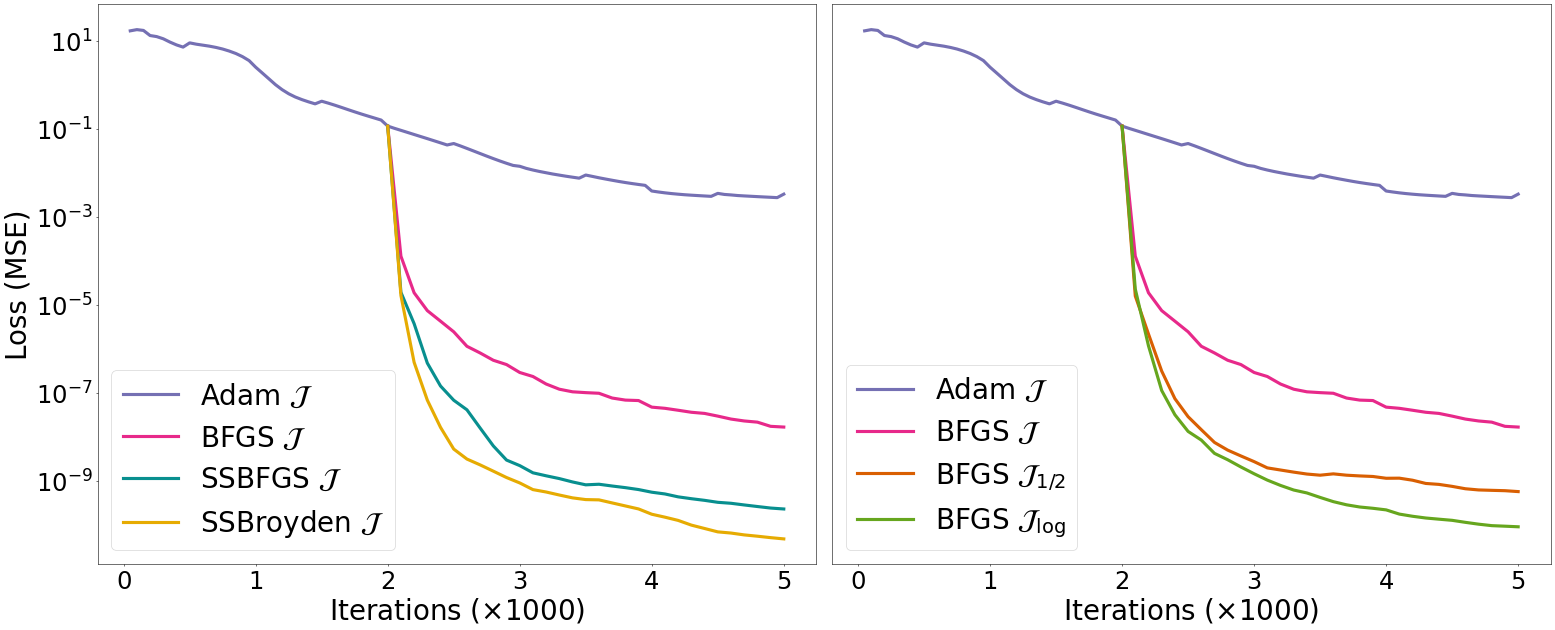}
    \caption{Evolution of the loss function with training iterations for the current-free Grad-Shafranov equation. In the left panel, results are obtained with the four optimization algorithms considered in section \ref{sec:optimization_algorithms}. In all cases, the standard MSE loss function $\mathcal{J}$ is used for training and the Adam optimizer is employed for the initial training phase. In the right panel, results are obtained with the loss function modifications considered in section \ref{sec:loss_function}. In all cases, the Adam optimizer is employed for the initial training phase and the BFGS algorithm for the quasi-Newton stage. Note that the plot shows the MSE loss $\mathcal{J}$ and not $\mathcal{J}_{1/2}$ or $\mathcal{J}_{\mathrm{log}}$ that were used during training.}
    \label{fig:loss_comparison}
\end{figure}

The left panel of figure \ref{fig:loss_comparison} shows the evolution of the loss function with the number of iterations for the four optimizers considered.
In all cases, we use Adam for the initial training phase in order to avoid possible saddle points and get closer to a global minimum before accelerating convergence with a quasi-Newton method.
The impact on convergence of the quasi-Newton formulae is glaring when compared to Adam.
BFGS achieves a loss function that is six orders of magnitude smaller than Adam, which is reduced by a further two and three orders of magnitude by its modifications SSBFGS and SSBroyden respectively.
This is further reflected in the absolute and relative errors of $\mathcal{P}$ and the magnetic field components (which depend on the first derivatives of $\mathcal{P}$), as can be seen in table \ref{tab:errors_LGS}.

The established interpretation of this phenomenon attributes it to a poorly scaled loss function $\mathcal{J}$, which results in an ill-conditioned Hessian matrix $\mathrm{hess}(\mathcal{J})$ close to the minimum \cite{OrenLuenberger, GKX19, WTP21}.
This characteristic is intrinsic to PINN configurations and is associated with formulating the loss function based on a differential operator \cite{dRBMdB23, RLFLU24}.

\begin{figure}
    \centering
    \includegraphics[width=.999\textwidth]{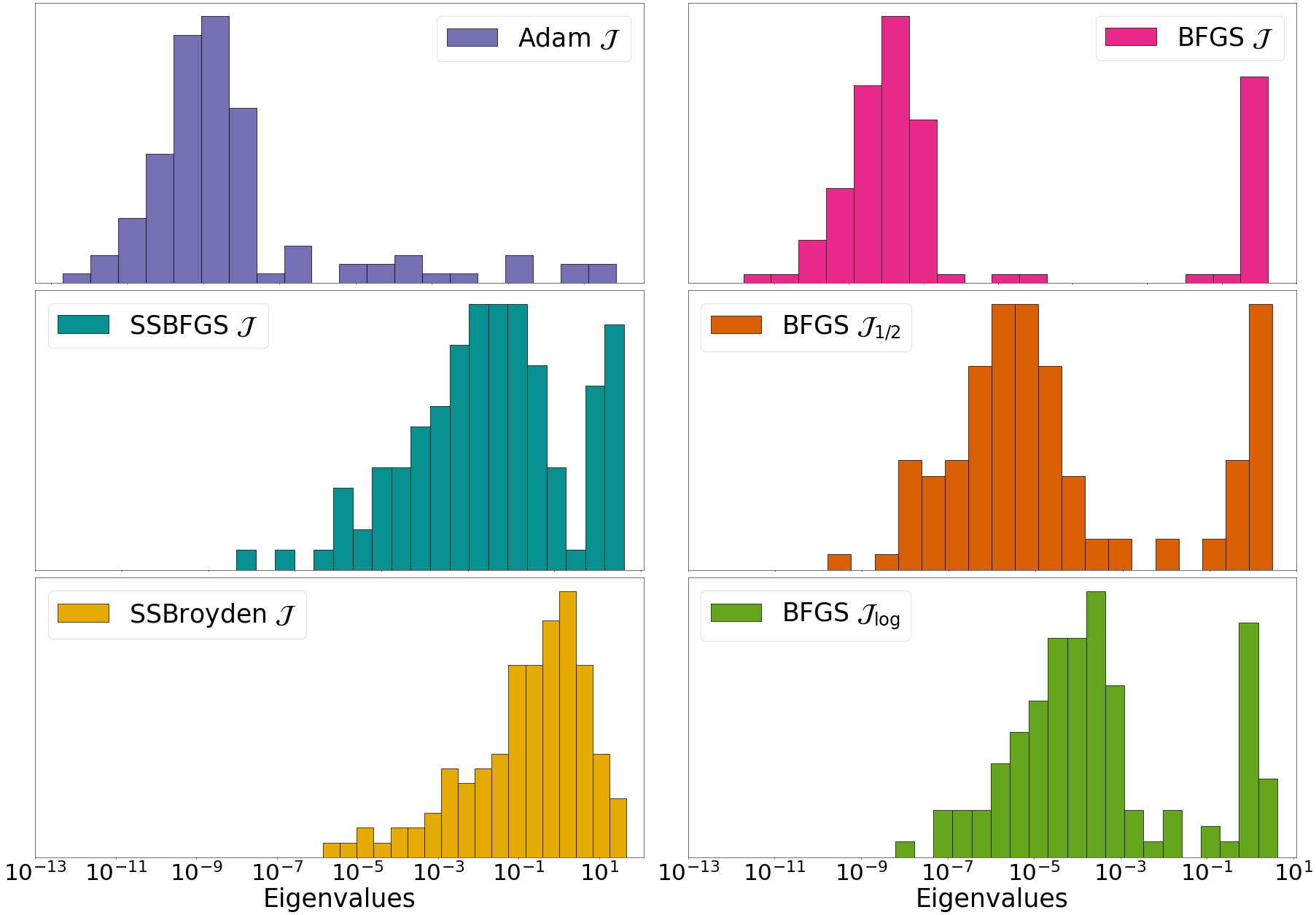}
    \caption{Eigenvalue spectra of the Hessian matrix $\mathrm{hess} (\mathcal{J}(\bm{z}))$ in the $\bm{z}$-space close to the minimum ($\mathcal{J} \sim 10^{-13}$) for the cases discussed in section \ref{sec:current_free_gs}. The upper left panel corresponds to no preconditioning ($H_k~=~I, \bm{z}~=~\bm{\Theta})$. The upper right panel corresponds to the usual BFGS optimization algorithm and MSE loss function. The middle and lower left panels correspond to the BFGS formula modifications while the middle and lower right panels correspond to the loss function modifications.}
    \label{fig:eigenvalue_spectrum}
\end{figure}

One approach to grasp this phenomenon is through the examination of the spectrum of the Hessian matrix. As depicted in the upper left panel of figure \ref{fig:eigenvalue_spectrum}, $\mathrm{hess}(\mathcal{J})$ demonstrates a broad eigenvalue spectrum, with numerous eigenvalues closely approaching zero and some outliers displaying larger magnitudes. Additionally, the condition number, represented by
\begin{equation}
\kappa = \frac{\lambda_{max}}{\lambda_{min}},
\end{equation}
where $\lambda_{max}$ and $\lambda_{min}$ denote the eigenvalues of the largest and smallest magnitude respectively, is notably large (on the order of $\kappa \sim 10^{12}$).
These characteristics confirm the ill-conditioning of $\mathrm{hess}(\mathcal{J})$,  indicating that certain directions within the trainable parameter space $\bm{\Theta}$ lead to significantly larger changes in the loss function compared to others.
In the conventional analogy of ``descending a mountain'', this scenario aligns with encountering long, extended valleys where certain directions exhibit steep gradients while others remain relatively flat.

Gradient descent techniques like Adam proceed by advancing along the steepest direction, leading to a perpetual zig-zagging along the valley and making minimal progress towards the minimum.
Quasi-Newton methods outperform Adam by incorporating knowledge of the local curvature of $\mathcal{J}$ via the Hessian, thus identifying superior descent directions.
However, findings from computational experiments in \cite{Brodlie1977} and \cite{ShannoPhua} reveal that the efficacy of the standard BFGS algorithm might still suffer due to ill-conditioning. They argue that it could be challenging for line search methods to determine a suitable step size 
$\alpha_k$, and suggest the use of self-scaled methods as a countermeasure.

Another way of understanding the influence of quasi-Newton methods, as pointed out in \cite{dRBMdB23, RLFLU24}, is through the preconditioning of the loss function.
At every iteration step, the inverse Hessian approximation acts as a preconditioner that maps the parameter space $\bm{\Theta}$, where the Hessian is ill-conditioned, to a new space $\bm{z}= H_k^{-1/2} \bm{\Theta}$, where the conditioning is much better. The rate of convergence of the different quasi-Newton methods will be strongly affected by the conditioning of the Hessian in this new space. 
To observe this effect mathematically, consider that representing
equation \eqref{eq:next_iterate} in the $\bm{z}$ space corresponds to the usual formula of a gradient-descent method
\begin{equation}
    \bm{z}_{k+1} = \bm{z}_k - \alpha_k \nabla\mathcal{J}(\bm{z}_k).
\end{equation}

However, the landscape of the loss function in this space is considerably more uniform (without long valleys) compared to the $\bm{\Theta}$ space. 
A step along the direction of the gradient genuinely advances towards the minimum.   
Meanwhile, as we approach the minimum, the Hessian of the new space $\bm{z}$
\begin{equation}\label{eq:hessian_z_space}
\mathrm{hess}\left(\mathcal{J}\left(\bm{z}\right)\right) = H_k^{1/2}\mathrm{hess}\left(\mathcal{J}\left(\bm{\Theta}\right)\right)H_k^{1/2},
\end{equation}
might also start to be ill-conditioned and the rate of convergence would, eventually, deteriorate.
This depends crucially on the updating formula of $H_k$, which explains the improved performance of the modifications of the standard BFGS formula that we investigate here. 

Please note that while the exact Newton's formula theoretically promises perfect conditioning, ensuring uniform gradients in all directions, its practical implementation poses numerical challenges.  
This is due to the necessity of solving an ill-conditioned linear system of equations to determine the Newton descent direction. Typically, such ill-conditioning arises from slight variations in matrix coefficients, particularly in the Hessian, resulting in significant deviations in the solution accuracy (see for example \cite{DOUGLAS2016941}).

To appreciate the effect of the Hessian conditioning, the upper right, middle left and lower left panels in figure \ref{fig:eigenvalue_spectrum} show the spectra of the Hessian expressed in the $\bm{z}$ space for each optimization algorithm. Each of these histograms is computed by letting the PINN to be trained additional iterations with a fixed training set until we arrive to similar values of $\mathcal{J}$, which we set to be very low ($\mathcal{J} \sim 10^{-13}$) in order to be close to the minimum. As we can observe, the standard BFGS algorithm gives an ill-conditioned Hessian in the $\bm{z}$ space close to the minimum, whereas with the BFGS modifications we obtain better-conditioned matrices. This is reflected in the number of iterations needed for each algorithm to arrive at the value of loss prescribed above: the BFGS algorithm needed $\sim 40000$ iterations, whereas the BFGS modifications only needed $\sim 1000$.

\subsubsection{Impact of the loss function.}

In a similar light, we can examine the impact of utilizing the modified versions of the loss function discussed in section \ref{sec:loss_function}. To this end, we train 
 three identical networks with different loss functions, namely $\mathcal{J}, \mathcal{J}_{1/2}, \mathcal{J}_{\log}$.
We use the standard BFGS algorithm in order to isolate the impact of the loss function modifications from that of the BFGS modifications.

When $\mathcal{J} > 1$, the loss function modifications could decelerate convergence because of the decreased slope of $\mathcal{J}_g$.
To prevent this, we utilize $\mathcal{J}_g$ exclusively during the BFGS phase of training, while maintaining the use of $\mathcal{J}$ during the Adam stage.
The right panel of figure \ref{fig:loss_comparison}
illustrates the effect of the loss function modifications on convergence.
Please observe that, for comparison purposes, we represent the standard MSE loss $\mathcal{J}$ in the plot, while the actual training has been conducted using the corresponding $\mathcal{J}_g$.
We achieve an improvement of roughly two orders of magnitude by evaluating the loss through a monotonic function $g$, everything else being equal.
$\mathcal{J}_{\log}$ converges slightly faster than $\mathcal{J}_{1/2}$ because its slope is steeper close to the minimum.
This enhancement becomes more evident when examining the relative error norms of the PDE solution $\mathcal{P}$ and its derivatives, as presented in table \ref{tab:errors_LGS}.

As in the previous section, one can attribute the improved convergence to the better conditioning of the Hessian matrix.
We employ the same optimization algorithm in all cases.
However, since we minimize different loss functions, the inverse Hessian approximations vary at each iteration. Consequently, the preconditioned Hessian $\mathrm{hess}(\mathcal{J}_g(\bm{z}))$ may exhibit different condition numbers $\kappa$ depending on the selection of $g$.
Indeed, by analyzing the eigenvalue spectra in the middle and lower right panels of figure \ref{fig:eigenvalue_spectrum}, 
it is apparent that $\mathcal{J}_{\log}$ exhibits a smaller condition number compared to $\mathcal{J}$, with the majority of its eigenvalues concentrated at higher values and more eigenvalues clustered around unity.
All these features suggest a better-conditioned Hessian matrix and explain its superior performance. Similar observations hold for $\mathcal{J}_{1/2}$.

Changes to the loss function can be implemented combined with the selection of a better optimization algorithm. 
For the range of choices that we have explored, the latter seem to have a more pronounced effect on convergence.
Using a combination of the best choices in each case can lead to overall better results.
In practice, however, it is much simpler to change the loss function than to modify an existing optimizer or to develop a new one.


\begin{table}[H]
    \centering
    \small
    \caption{Relative error (equation \eqref{eq:error_norm}) of the flux function $\mathcal{P}$ and the magnetic field components $B_r, B_\theta$ for the current-free Grad-Shafranov equation, employing different combinations of optimizer/loss function. 
    The errors are averaged over multiple training runs. These runs employ different random initializations of the trainable parameters to ensure robustness. The results are presented in the format (mean ± standard deviation), giving a clear indication of both the average error and the variability across the different trials.}
    \begin{tabular}{l l c c c}
        \hline
        Optimizer & Loss & $E^{(2)}_{\mathcal{P}}$ & $E^{(2)}_{B_r}$ & $E^{(2)}_{B_\theta}$  \\
        \hline
        BFGS & $\mathcal{J}$ & $(2.8 \pm 1.3) \times 10^{-6}$ & $(2.1 \pm 1.2) \times 10^{-5}$ & $(9 \pm 2) \times 10^{-6}$ \\ 
        BFGS & $\mathcal{J}_{1/2}$ & $(3.6 \pm 0.9) \times 10^{-7}$ & $(4 \pm 2) \times 10^{-6}$ & $(1.1 \pm 0.3) \times 10^{-6}$ \\ 
        BFGS & $\mathcal{J}_{\mathrm{log}}$ & $(2.3 \pm 0.9) \times 10^{-7}$ & $(2.4 \pm 0.9) \times 10^{-6}$ & $(7 \pm 3) \times 10^{-7}$ \\ 
        SSBFGS & $\mathcal{J}$ & $(1.9 \pm 0.4) \times 10^{-7}$ & $(2.3 \pm 0.7) \times 10^{-6}$ & $(4.5 \pm 0.6) \times 10^{-7}$ \\ 
        SSBroyden & $\mathcal{J}$ & $(1.0 \pm 0.2) \times 10^{-7}$ & $(1.0 \pm 0.3) \times 10^{-6}$ & $(2.5 \pm 0.7) \times 10^{-7}$ \\
        \hline
    \end{tabular}
    \label{tab:errors_LGS}
\end{table}

\subsection{Non-linear force-free solutions with higher order multipoles.}\label{sssec:force_free}

\begin{figure}[H]
    \centering
    \includegraphics[width=.7\textwidth]{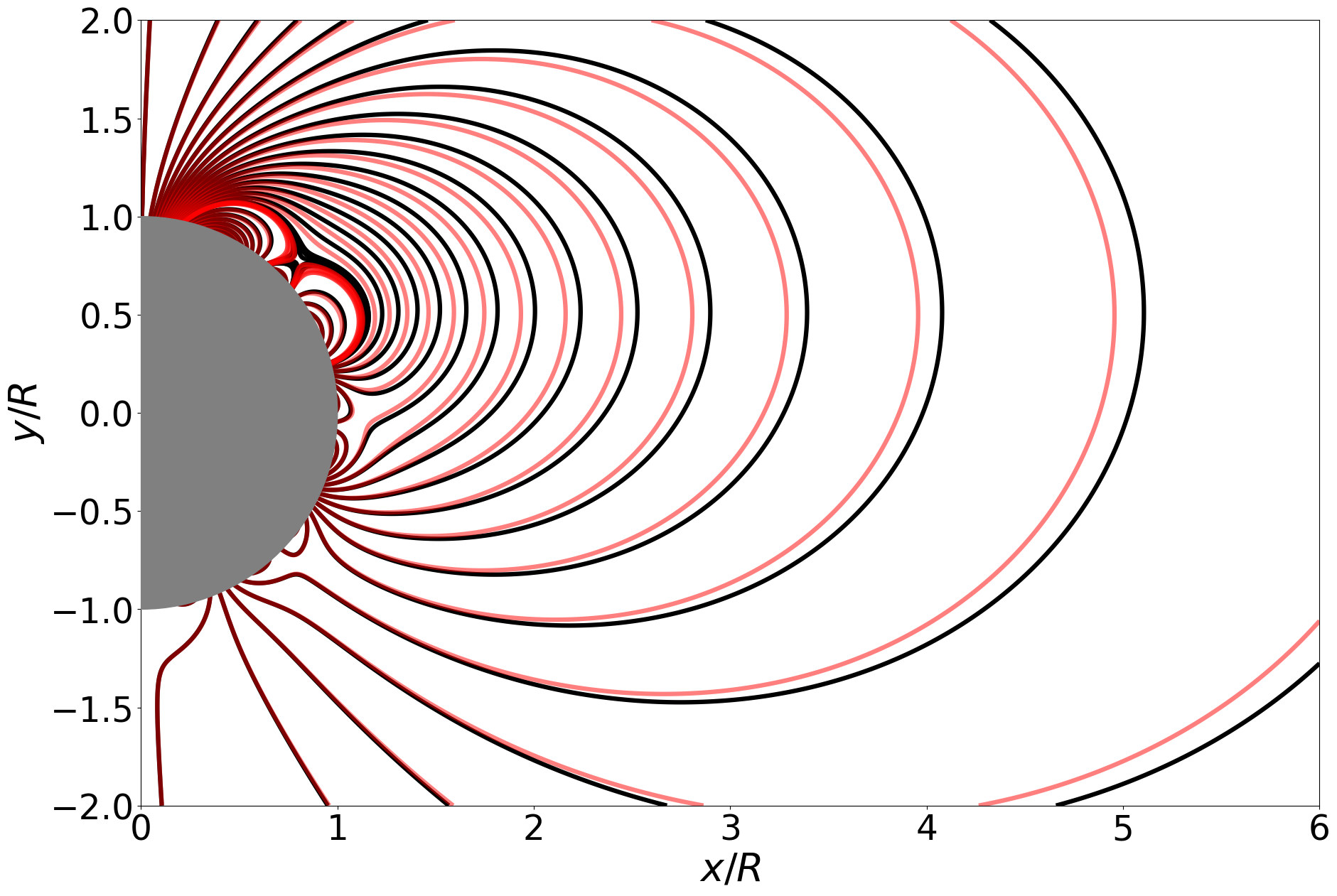}
    \caption{Magnetic field lines obtained for the force-free case (black). The field lines corresponding to the current-free case (red) are also plotted for comparison.}
\label{fig:magnetic_field_lines_8_multipoles_force_free}
\end{figure}

Up to now, our emphasis has been on a relatively simple, linear problem as to illustrate how enhancements in the optimization process can significantly improve the solution accuracy.
Moving forward, we now show that our results generalize nicely to more complex solutions, in particular, we consider the Non-Linear Grad-Shafranov (NLGS) equation.
We introduce a toroidal function similar to the one chosen in \cite{TanerForceFree} but generalized for negative values of $\mathcal{P}$. This is:
\begin{equation}
    \mathcal{T}\left(\mathcal{P}\right) = 
    \begin{cases}
         s \left(\left|\mathcal{P}\right| - \mathcal{P}_c\right)^\sigma &\mathrm{ if } \left | \mathcal{P} \right| > \mathcal{P}_c \\
        0 &\mathrm{ if } \left | \mathcal{P} \right| < \mathcal{P}_c,
    \end{cases}
\end{equation}
where $(s,\mathcal{P}_c, \sigma)$ are parameters that control the relative strength of the toroidal and the poloidal magnetic fields, the extent of the current-filled region, and the degree of non-linearity of the model, respectively. 
At the surface of the star, the boundary condition consists of eight multipoles, so that $b_{l \le 8} \ne 0$ in equation \eqref{eq:surface_bc}. An example solution is shown in figure \ref{fig:magnetic_field_lines_8_multipoles_force_free}.

The presence of non-linearity and a highly multipolar structure significantly increase the solution's complexity. Consequently, larger networks with a larger number of trainable parameters are necessary to attain comparable precision.
Nevertheless, because of the efficient optimization, we manage to keep the network size rather small even for this problem.
Table \ref{tab:training_params} contains the training and architecture hyperparameters.

\begin{figure}
    \centering
    \includegraphics[width=0.999\linewidth]{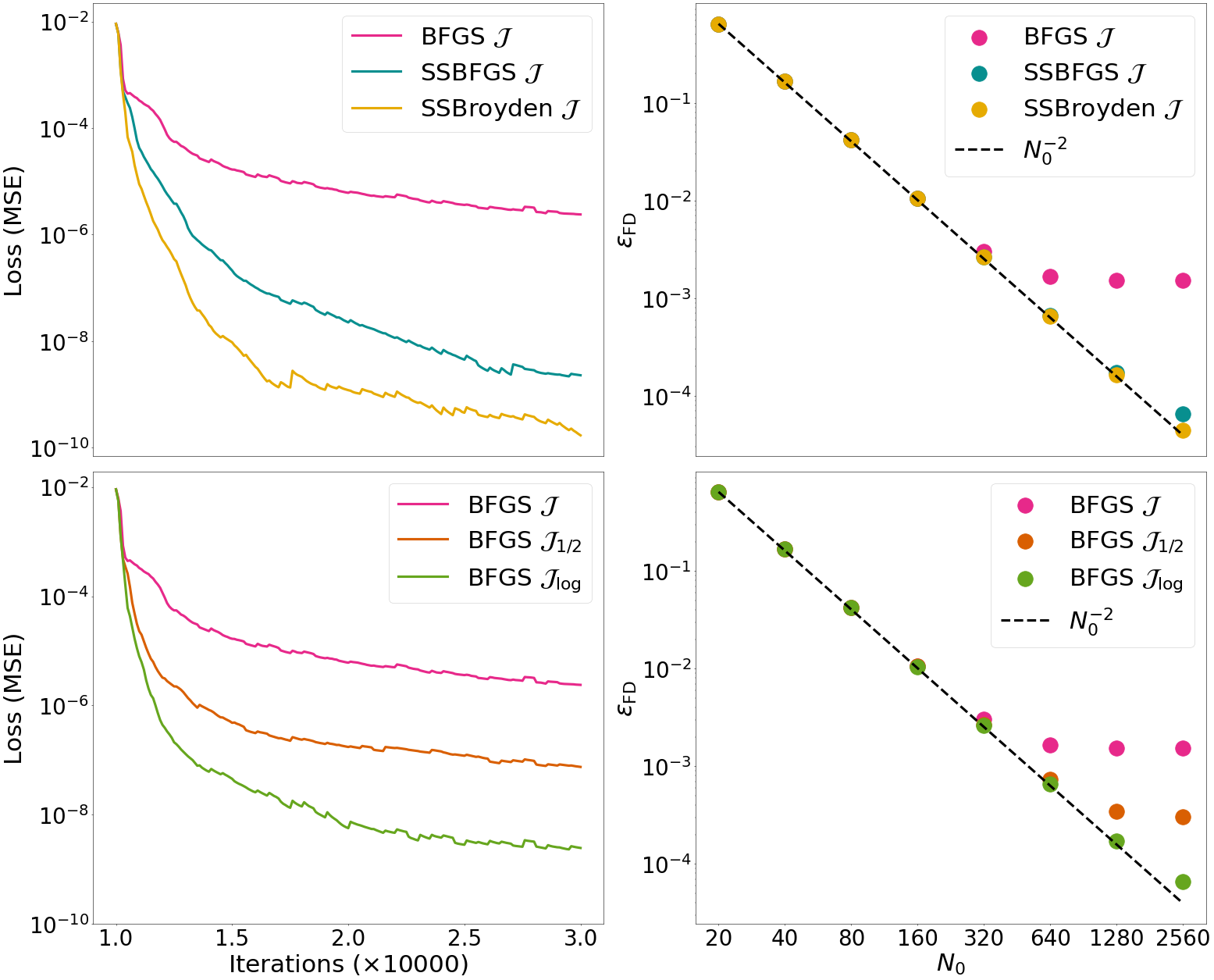}
    \caption{Evolution of the loss function with training iterations (left panels) and $L_2$ norm of the discretized PDE (see \ref{app:errornorms}) as a function of the grid resolution $N_0$ (right panels) for the non-linear Grad-Shafranov equation. The upper panels correspond to the optimizer modifications while the lower panels correspond to the loss function modifications.}
    \label{fig:force_free_GS_loss_and_error}
\end{figure}

Figure \ref{fig:force_free_GS_loss_and_error} shows the performance of the BFGS modifications (upper panels) and of the loss function modifications (lower panels) on this problem.
The left panels display the evolution of the loss function as the number of iterations increases, while the right panels present the
$L_2$-error of the discretized PDE as a function of the number of grid points in the test grid. The minimum of this error is used to estimate the overall error $\epsilon_{\mathrm{NN}}$, as detailed in \ref{app:errornorms}.
The improvement in convergence and precision is substantial for this more complex problem, similar to the simpler problem of section \ref{sec:current_free_gs}. Table \ref{tab:errors_NLGS} shows the specific values of $\epsilon_{\mathrm{NN}}$ for the different choices of optimizer and loss function.

The PINN error when the standard BFGS algorithm is used for training is about $\epsilon_{\mathrm{NN}} \sim 10^{-3}$.
SSBFGS and SSBroyden achieve much lower errors, of the order of $10^{-5}$.
Specifically, $\epsilon_{\mathrm{NN}}$ begins to surpass $\epsilon_{\mathrm{FD}}$ at an approximate resolution of $4500 \times 4500$ (not depicted in the figure), near the memory limit of the machine employed for generating the results in this paper.
Note that, since our discretization scheme is second order, the absolute error of the PDE is $\epsilon_{\mathrm{NN}}^2$.
In summary, aiming for a comparable precision level ($<10^{-4}$) with a second-order finite difference scheme would require a grid comprising at least $1000 \times 1000$ points for this 2D problem. In higher dimensions, the advantages of PINNs would become even more pronounced. 

\begin{table}
     \centering
     \small
    \caption{PINN approximation error $\epsilon_{\mathrm{NN}}$ (equation \eqref{eq:eNN_definition}) for the non-linear Grad-Shafranov equation. The errors are averaged over multiple training runs. These runs employ different random initializations of the trainable parameters to ensure robustness. The results are presented in the format (mean ± standard deviation), giving a clear indication of both the average error and the variability across the different trials.}
     \begin{tabular}{l l c}
         \hline
         Optimizer & Loss &  $\epsilon_{\mathrm{NN}}$  \\ 
         \hline
         BFGS & $\mathcal{J}$ & $(1.8 \pm 0.2 ) \times 10^{-3}$  \\ 
         BFGS & $\mathcal{J}_{1/2}$ & $(4.0 \pm 0.7) \times 10^{-4}$  \\ 
         BFGS & $\mathcal{J}_{\mathrm{log}}$ & $(7 \pm 4) \times 10^{-5}$ \\
         SSBFGS & $\mathcal{J}$ & $(4.0 \pm 1.2 ) \times 10^{-5}$ \\ 
         SSBroyden & $\mathcal{J}$ & $(1.5 \pm 0.2 ) \times 10^{-5}$  \\ 
         \hline
     \end{tabular}
     \label{tab:errors_NLGS}
 \end{table}

\subsection{Parameter study}\label{sec:parameter_study}

\begin{figure}[H]
    \centering
\includegraphics[width=0.999\linewidth]{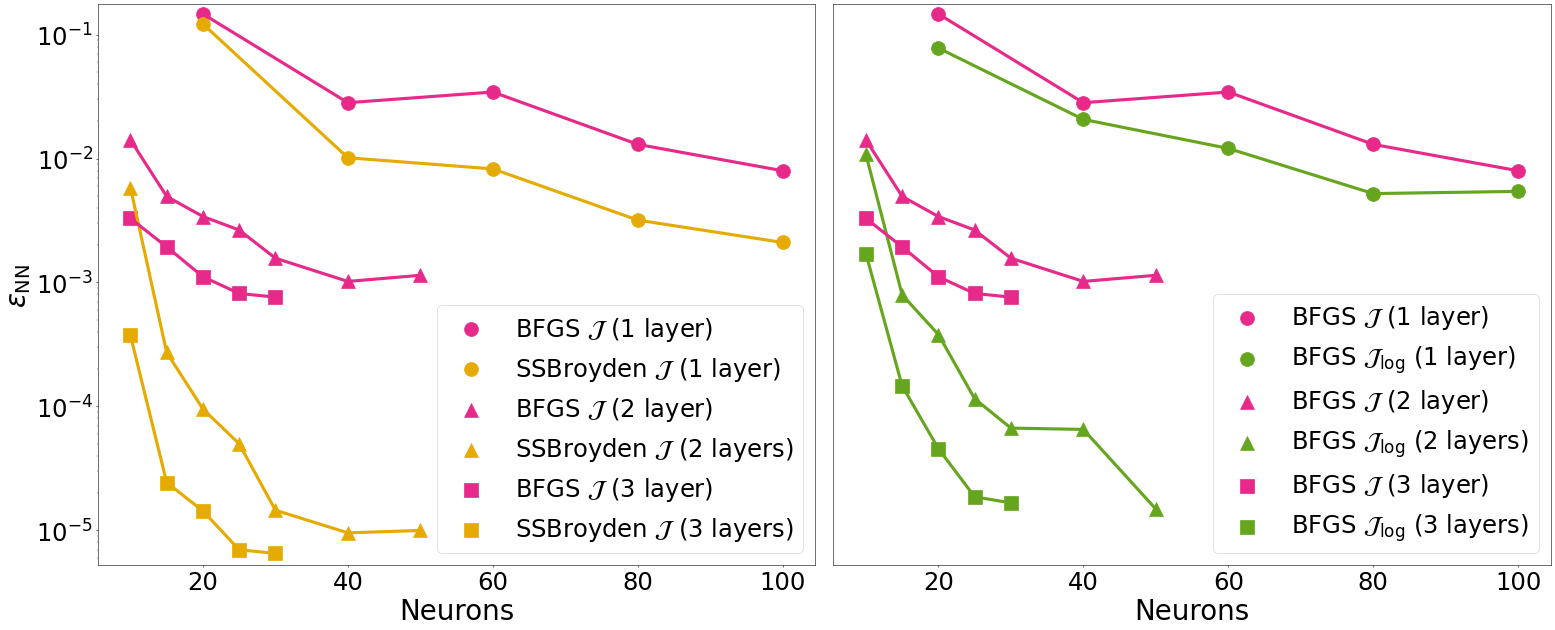}
    \caption{Approximation error $\epsilon_{\mathrm{NN}}$ for different depths as a function of the neurons in each layer. Left panel: comparison between using BFGS and $\mathcal{J}$ and SSBroyden with $\mathcal{J}$. Right panel: comparison between using BFGS with $\mathcal{J}$ and BFGS with $\mathcal{J}_{\mathrm{log}}$ .}
    \label{fig:hyperparameter_study}
\end{figure}

The impact of the modifications introduced before may depend on the size of the network. Hence, we now focus on a hyperparameter study, where we explore how the accuracy of the PINN approximation varies with the number of neurons per layer or the number of layers. For each model, 
we keep fixed all choices regarding the number of training points, the number of iterations, etc. Figure \ref{fig:hyperparameter_study} shows the values of $\epsilon_{\mathrm{NN}}$ as a function of the neurons in each layer, obtained for the SSBroyden algorithm with $\mathcal{J}$ and for the BFGS algorithm with $\mathcal{J}_{\mathrm{log}}$ for different depths (that is, varying the number of hidden layers). The results obtained with BFGS and $\mathcal{J}$ are also plotted for comparison purposes. 

For simpler networks consisting of just one layer, these adjustments have a relatively minor effect. However, for networks with 2-3 layers, there is a significant boost in accuracy when a certain minimum number of neurons is employed. While simply increasing the number of parameters 
leads to a gradual reduction in error, our proposed enhancements lead to a much faster decline. 
This underscores the significance of refining the optimization process over indiscriminately enlarging the network size. 
The right panel of figure \ref{fig:hyperparameter_study} demonstrates that the impact of the optimizer outweighs slightly the effect of redefining the loss function. However, it is worth noting that the latter adjustment involves simply changing a single line of code.

In summary, using networks comprising $2$ or $3$ hidden layers with around $30-40$ neurons each, we achieve highly precise results for this specific problem. In the following section, we will show that networks of similar dimensions suffice to yield results of similar accuracy across various physical applications governed by different equations.

\section{Other physics problems}\label{sec:other_problems}

Our analysis of the optimization process and how it can benefit from modifications of the optimizer or the loss function is not problem specific and is not tailored to neutron star magnetospheres.
To showcase its broad effectiveness, we now discuss solutions for diverse PDEs across various fields and applications.
In this section, we cover cases with higher order derivatives, more dimensions, various degrees of non-linearity, time-dependence, systems of equations etc.

For each case, we provide a concise description of each PDE along with the context in which it is applied. Then we specify how we formulate each problem in accordance with the notation of section \ref{sec:summary}.
The architecture and training hyperparameters can be found in table \ref{tab:training_params2}.

\begin{table}[]
    \centering
    \small
    \caption{Architecture and training hyperparameters for various physical applications considered in section \ref{sec:other_problems}. In all cases, we use a $\tanh$ activation function for the hidden layers. The \textit{Neurons} column refers to neurons per hidden layer. The \textit{Adam it.} column refers to the number of iterations where the Adam optimizer is used before switching to a quasi-Newton method. The \textit{Batch size} column refers to the number of points sampling the domain for each training set. The training set changes every 500 iterations in order to sample as many points as possible. The \textit{Iterations} column denote the number of Quasi-Newton iterations.}
    \begin{tabular}{l c c c c c c}
        \hline 
        PDE & Layers & Neurons & Iterations & Adam it. & Batch size & Domain 
        \\   & & & (x1000) & (x1000)& (x1000) &
        \\ [0.5ex]
        \hline 
        2DH $(1,4)$ & 2 & 20 & 20 & 5 & 10 & $[-1,1] \times [-1,1]$ \\
        2DH $(6,6)$ & 3 & 30 & 50 & 5 & 10 & $[-1,1] \times [-1,1]$ \\
        NLP & 2 & 30 & 20 & 10 & 8 &  $[-1,1] \times [-1,1]$ \\ [0.5ex]
        NLS & 2 & 40 & 20 & 10 & 10 & $\left[0,\pi/2 \right] \times [-15,15]$ \\ [0.5ex]
        KdV & 3 & 30 & 20 & 10 & 15& $\left[0,5 \right] \times [0,20]$  \\ [0.5ex]
         1DB & 3 & 20 & 10 & 5 & 10 & $\left[0,1 \right] \times \left[-1,1 \right]$ \\
         AC & 3 & 30 & 20 & 5 & 10 & $\left[0,1 \right] \times \left[-1,1 \right]$ \\
        3DNS & 2 & 40 & 20 & 10 & 10 & $[0,1] \times [-1,1]^3 $ \\ [0.5ex]
         LDC & 6 & 20 & 20 & 0 & 25 & $[0,1] \times [0,1]$ \\
        \hline
    \end{tabular}
    \label{tab:training_params2}
\end{table}

In all cases, we use the Adam optimizer and the standard MSE loss $\mathcal{J}$ for the initial training phase.
After that, we train using either BFGS with $\mathcal{J}_{\log}$ or using SSBroyden with $\mathcal{J}$, which were the most competitive combinations of modifications in section \ref{sec:ns_magnetosphere}.

For each problem, we show the evolution of the loss function with the number of iterations and the evolution of the relative $L_2$ error with respect to some reference solution, reaffirming that the findings outlined in section \ref{sec:ns_magnetosphere} hold universally. In addition, we present a comparison of the PINN prediction against the reference solution (the exact solution when available) and the distribution of errors in the domain.

\paragraph{\textbf{2D Helmholtz equation (2DH)}}

We begin by examining the 2D Helmholtz equation problem as outlined in  \cite{ANAGNOSTOPOULOS2024116805,ASTSK2024}:
\begin{equation}\label{eq:2DH}
    \nabla^2 u + k^2 u - q(x,y) = 0,
\end{equation}
where $k$ is a constant, and the source term
\begin{equation}
    q(x,y) = -\sin (\pi a_1 x) \sin (\pi a_2 y) \left[\pi^2 \left(a_1^2 + a_2^2 \right) - k^2 \right],
\end{equation}
is chosen such as the solution to the problem is analytical  
$$u(x,y) = \sin \left(\pi a_1 x \right) \sin \left(\pi a_2 y \right),$$ where $a_1,a_2 \in \mathbb{Z}$. The computational domain is the square $\left[-1, 1 \right] \times \left[-1,1 \right] \in \mathbb{R}^2$.

Imposing periodic boundary conditions in the $x$ and $y$ directions, the PINN solution is
\begin{equation}
    u(x,y) = \mathcal{N}\left(\cos \pi x, \sin \pi x, \cos \pi y, \sin \pi y \right),
\end{equation}
where $\mathcal{N}$ is the output of the network. The loss function is then constructed using equation \eqref{eq:loss_function}, where $\mathcal{L} = \nabla^2$ and $G=-k^2 u + q(x,y)$. We consider a low wavenumber case with $a_1=1$, $a_2=4$ and $k=1$, as in \cite{ANAGNOSTOPOULOS2024116805}. We remark that, if we consider periodic boundary conditions, the constant $k$ should be chosen such that the solution of the homogeneous PDE with periodic boundary conditions is zero. It can be readily seen that the homogeneous solution is zero, provided that $k^2 \neq \pi^2 \left(n^2 + m^2 \right)$, where $n, m \in \mathbb{Z}$.

\begin{figure}[H]
    \centering
    \includegraphics[width=0.999\linewidth]{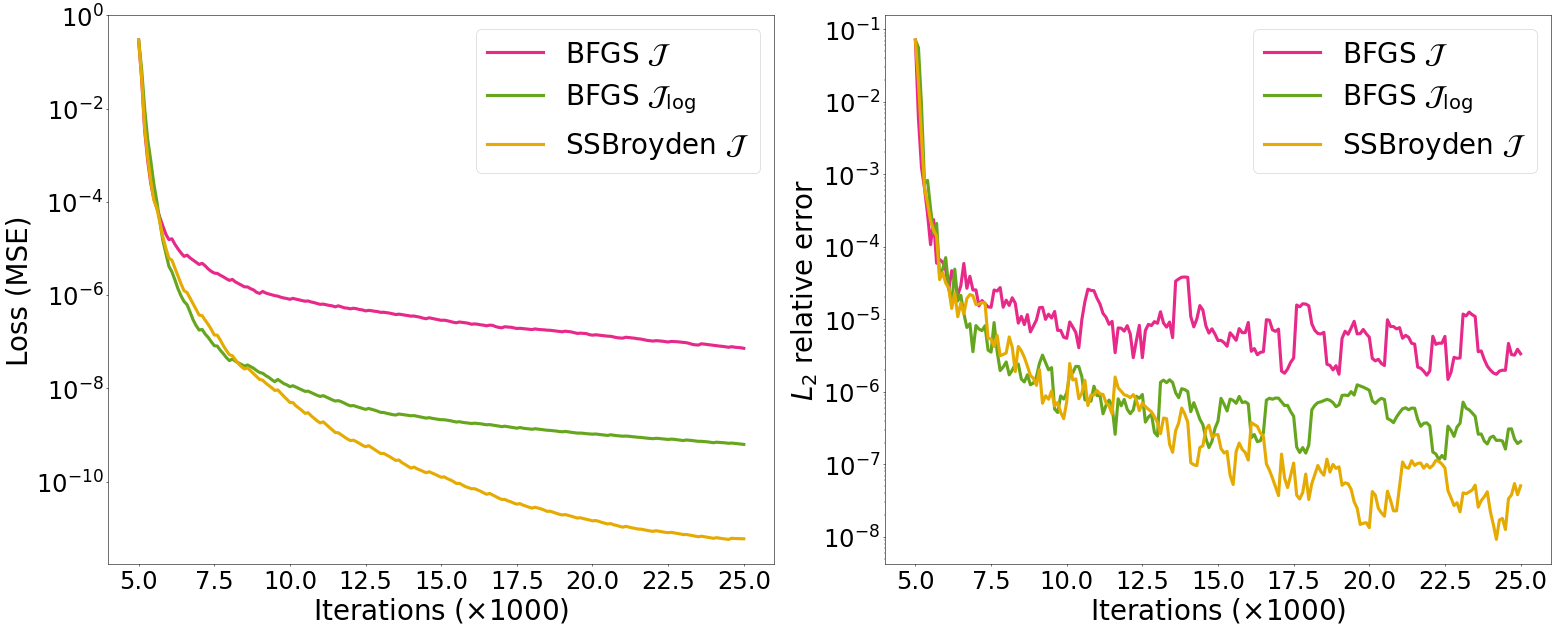}
    \caption{Convergence plots for the 2D Helmholtz equation with low wavenumber ($a_1=1$, $a_2=4$). Left panel: evolution of the loss function. Right panel: evolution of the relative $L_2$ error with respect to the reference solution on a fixed grid.}
    \label{fig:loss_helmholtz}
\end{figure}

\begin{figure}[H]
    \centering
     \includegraphics[width=0.999\linewidth]{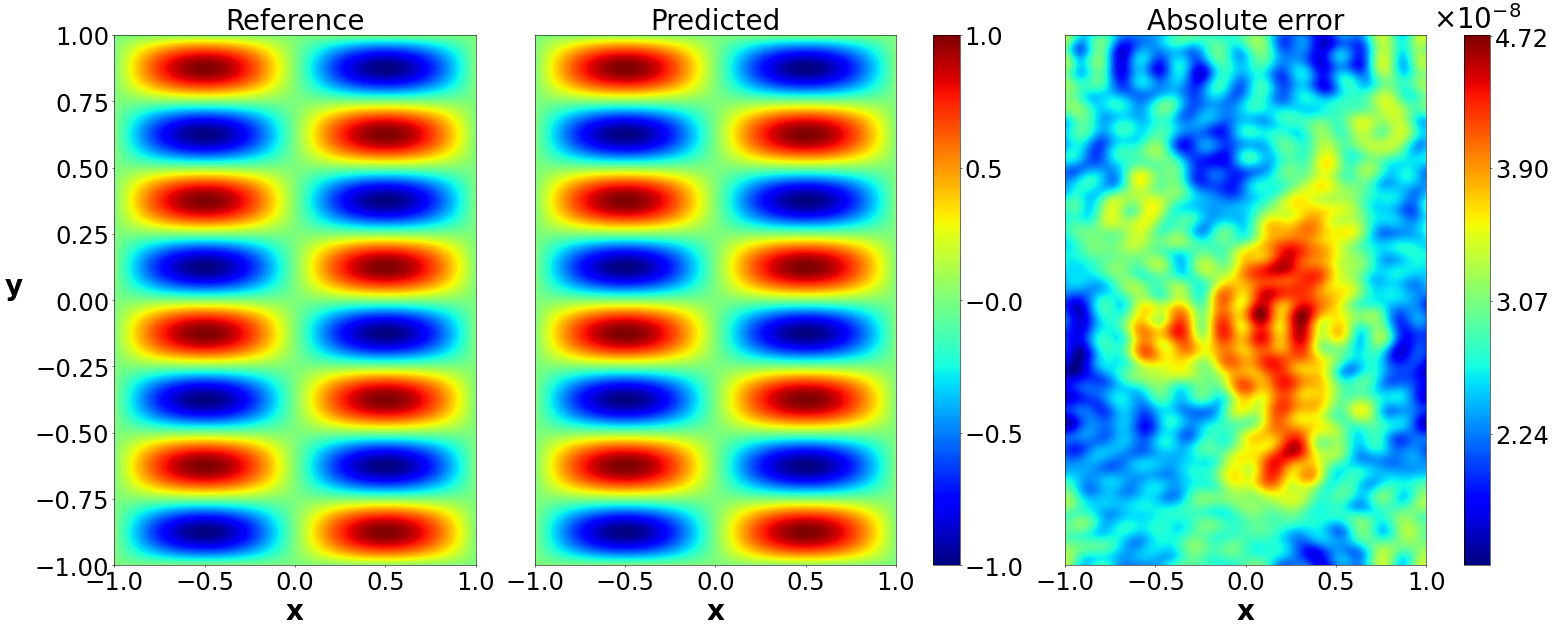}
    \caption{Solution comparison for the 2D Helmholtz equation with low wavenumber ($a_1=1$, $a_2=4$). Left panel: reference solution. Middle panel: PINN prediction obtained with the SSBroyden algorithm and the standard MSE loss. Right panel: absolute difference between the two. Note that the scale of the absolute error is $10^{-8}$.}
    \label{fig:helmholtz_solution}
\end{figure}

We refer to table \ref{tab:training_params2} for the specific set of hyperparameters chosen for this problem. The training set follows a random uniform distribution in the domain and is resampled every $500$ iterations. Figure \ref{fig:loss_helmholtz} shows the evolution of the loss function with iterations, employing the BFGS algorithm in conjunction with the MSE loss, the SSBroyden algorithm with the MSE loss, and the BFGS algorithm with the logarithm of the MSE loss $\mathcal{J}_{\mathrm{\log}}$. We notice again that the modifications of the BFGS algorithm and the loss function introduce a remarkable improvement in convergence, achieving a reduction of $\sim 2-3$ orders of magnitude at the end of the training process. This is reflected in the relative $L_2$ error, where our modifications achieved to reduce the error between $1$ and $2$ orders of magnitude.
Figure \ref{fig:helmholtz_solution} shows the analytical and the PINN solutions, together with the absolute difference between them.

\begin{figure}[H]
    \centering
     \includegraphics[width=0.999\linewidth]{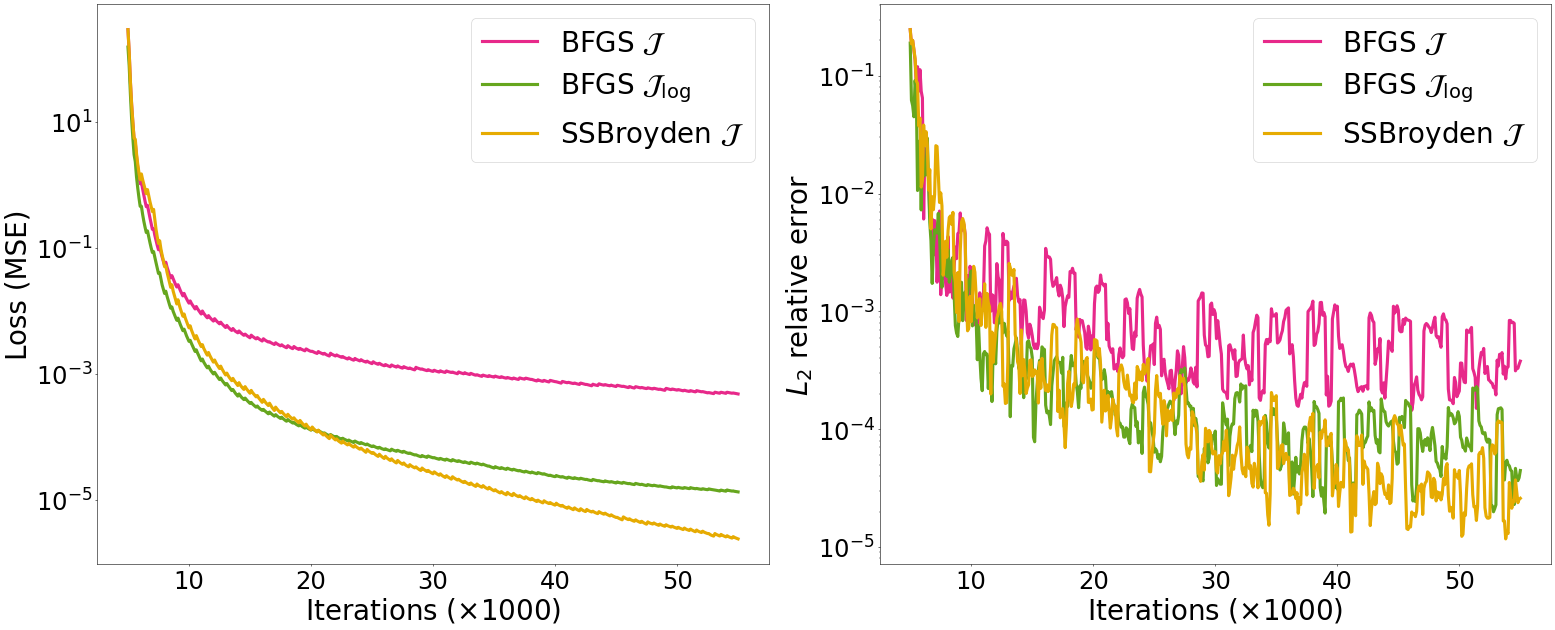}
    \caption{Convergence plots for the 2D Helmholtz equation with high wavenumber ($a_1 = a_2 = 6$). Left panel: evolution of the loss function. Right panel: evolution of the relative $L_2$ error with respect to the reference solution on a fixed grid.}\label{fig:loss_helmholtz_higher}
\end{figure}

\begin{figure}[H]
    \centering
 \includegraphics[width=.999\textwidth]{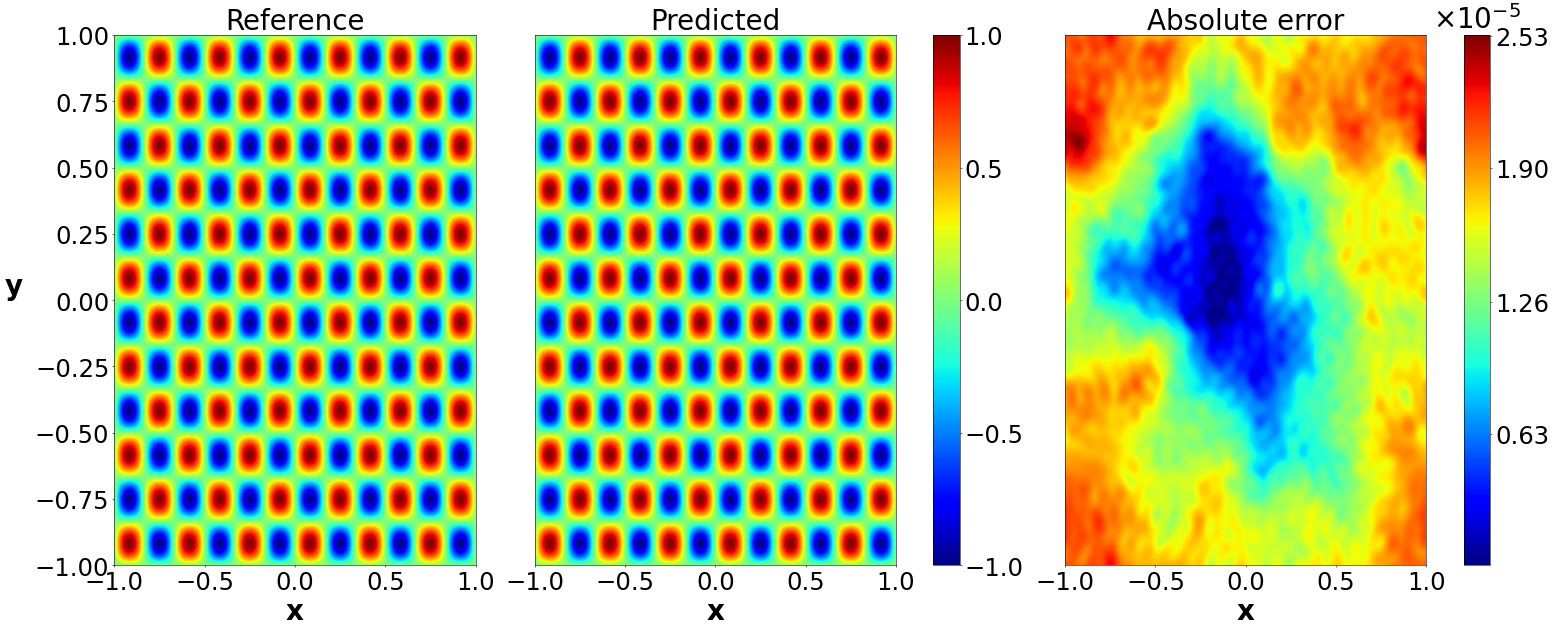} 
    \caption{Solution comparison for the 2D Helmholtz equation with high wavenumber ($a_1 = a_2 = 6$). Left panel: reference solution. Middle panel: PINN prediction obtained with the SSBroyden algorithm and the standard MSE loss. Right panel: absolute difference between the two.}
    \label{fig:helmholtz_solution_higher}
\end{figure}

We have also considered a more challenging case with higher wavenumber solution, where $a_1 = a_2 = 6$ and $k = 1$ as in \cite{ASTSK2024}. As expected, resolving smaller structures calls for increased network complexity (see table \ref{tab:training_params2}) to achieve accurate results, but the same improvements of the previous cases are observed in figure \ref{fig:loss_helmholtz_higher}. Once again, our modifications achieved to reduce the error significantly, obtaining a $L_2$ relative error of roughly one order of magnitude.
In figure \ref{fig:helmholtz_solution_higher} we show the analytical and the PINN solutions (employing the self-scaled Broyden algorithm), and the relative difference between them. Errors are summarized later in table \ref{tab:comparison_pinns}.

\paragraph{\textbf{Non-linear Poisson equation (NLP)}} 

We solve the Poisson equation with a non-linear (exponential) term, as considered in \cite{SharmaShankar2022}
\begin{equation}\label{eq:NLP}
    \nabla^2 \phi - e^{\phi} = r(x,y).
\end{equation}
If $r(x,y)= 0$, it is also called the Liouville equation in the context of differential geometry. 
It has application in various fields, such as hydrodynamics, to describe mean field vorticity in steady flows \cite{Caglioti1992,Chanillo1994} and Quantum Field Theory, in the Chern-Simons theory \cite{1990PhRvL..64.2230H, 1990PhRvL..64.2234J}.

To construct the loss function \eqref{eq:loss_function}, we can identify $\mathcal{L} = \nabla^2$ and $G = e^\phi + r(x,y)$.
The function $r(x,y)$ is chosen such that the function
\begin{equation}
    \phi(x,y) = 1 + \sin \left (k\pi x \right) \cos \left(k\pi y \right), 
\end{equation}
is a solution of the PDE, for some $k \in \mathbb{Z}$. 
We solve the problem in Cartesian coordinates $x_\alpha = (x, y)$ with Dirichlet boundary conditions
\begin{align}
    \phi (0,y) &= \phi(1,y) = 1, \\
    \phi (x,0) &= 1 + \sin \left(k\pi x \right), \\
    \phi(x,1) &= 1 + \sin \left(k\pi x \right)\cos \left(k \pi \right),
\end{align}
which can be hard-enforced through the following definitions
\begin{align}
    f_b(x,y) &= 1 + \left[1-y \left(1-\cos \left(k \pi \right)\right) \right] \sin \left(k\pi x \right), \\
    h_b(x,y) &= xy(1-x)(1-y).
 \end{align}
In \cite{SharmaShankar2022} they consider the simplest case with $k=1$.
We decided to raise this number to $k = 4$, in order to get a more pronounced oscillatory behavior, that would challenge our solver. 

\begin{figure}[H]
    \centering
     \includegraphics[width=0.999\linewidth]{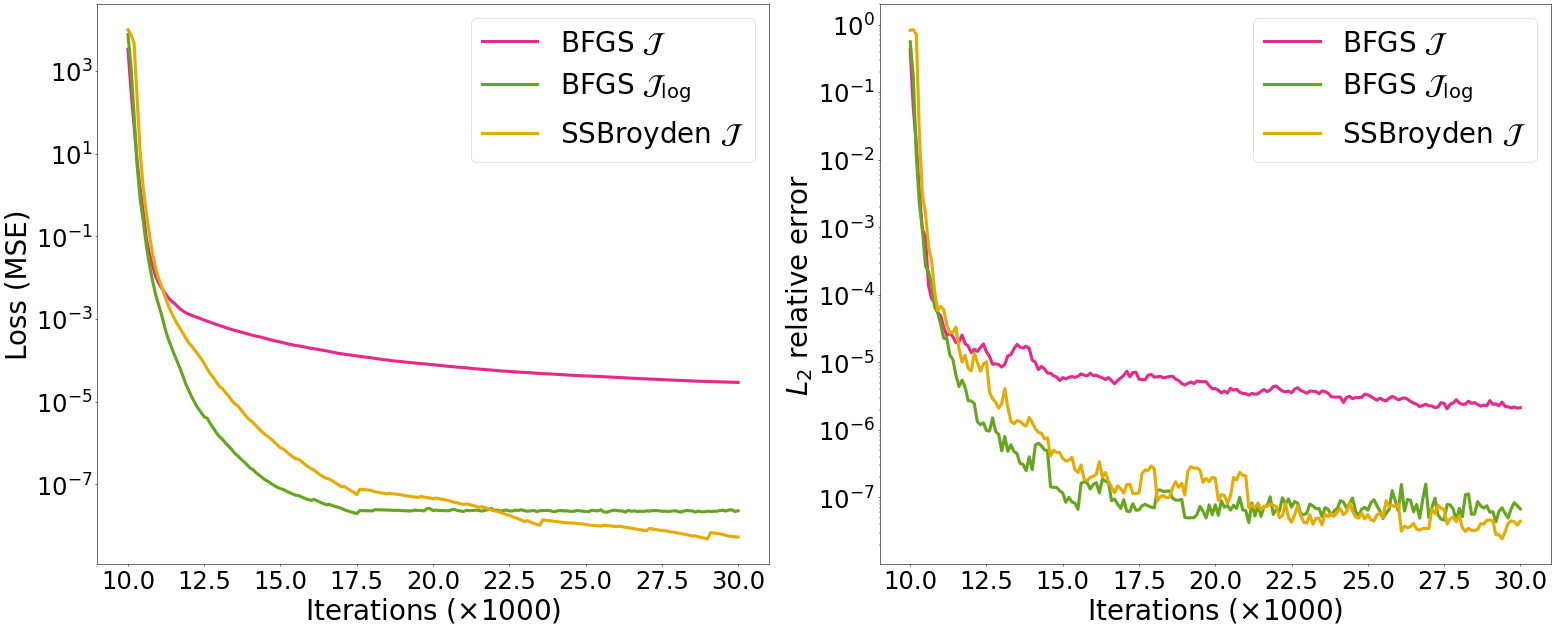}
    \caption{Convergence plots for the non-linear Poisson equation. Left panel: evolution of the loss function. Right panel: evolution of the relative $L_2$ error with respect to the reference solution on a fixed grid.}
    \label{fig:loss_NLP}
\end{figure}

\begin{figure}[H]
    \centering
    \includegraphics[width=.999\textwidth]{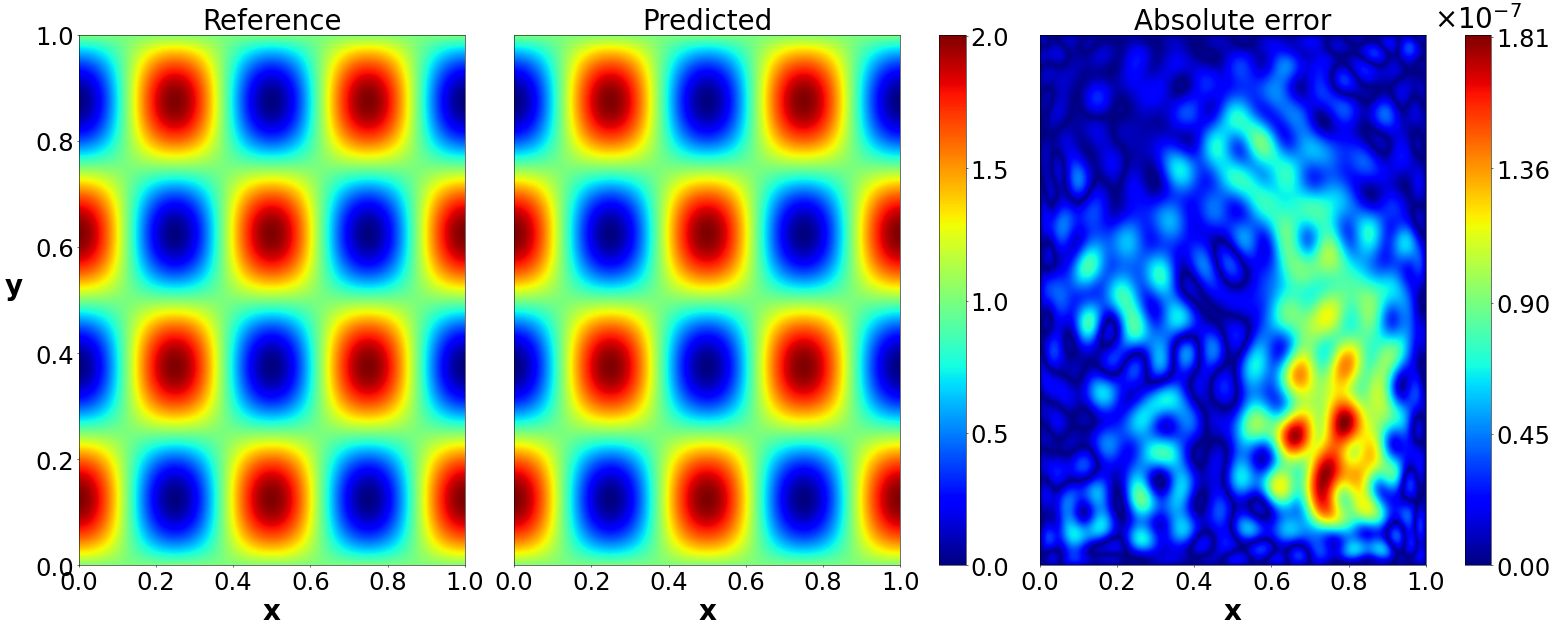} 
    \caption{Solution comparison for the non-linear Poisson equation. Left panel: reference solution. Middle panel: PINN prediction obtained with the SSBroyden algorithm and the standard MSE loss. Right panel: absolute difference between the two.}
    \label{fig:NLP_solution}
\end{figure}

Results for the loss function and the error norms are shown in figure \ref{fig:loss_NLP} and table \ref{tab:comparison_pinns} respectively. Figure \ref{fig:NLP_solution} shows the analytical solution, the PINN solution and their absolute difference, obtained with the self-scaled Broyden algorithm and the standard MSE loss.

\paragraph{\textbf{Non-linear Schrödinger equation (NLS)}} 

We solve the time-dependent Schrödinger equation in 1D, written in convenient units to avoid $\hbar$ prefactors as
\begin{equation}
    i \frac{\partial \Psi}{\partial t} = -\frac{1}{2}\frac{\partial^2 \Psi}{\partial x^2} + V \Psi, \label{eq:schrodinger_equation}
\end{equation}
where $i$ is the imaginary unit, and $V(\Psi) = - \left|\Psi \right|^2$ is a non-linear potential.
$\Psi$ is, in general, a function whose image lies in $\mathbb{C}$. 
Hence, if we denote $\Psi (x,t) \equiv u(x,t) + i v (x,t)$, we obtain the following non-linear coupled system of PDEs
\begin{align}
    &\frac{\partial v}{\partial t} - \frac{1}{2}\frac{\partial^2 u}{\partial x^2} - \left(u^2 + v^2 \right) u = 0, \label{eq:real_NLSE} \\ 
    &\frac{\partial u}{\partial t} + \frac{1}{2}\frac{\partial^2 v}{\partial x^2} + \left(u^2 + v^2 \right) v = 0. \label{eq:imaginary_NLSE}
\end{align}

The non-linear Schrödinger equation describes the dynamics of a non-linear wave packet in dispersive media. 
In the context of Bose-Einstein condensate, it is known as the Gross-Pitaevskii equation \cite{Gross1961,Pitaevskii1961}.
This equation is widely applicable in different physical scenarios, such as fluid mechanics, in order to model small-amplitude gravity waves \cite{HKAC18}, superconductivity and superfluidity \cite{GL1950,Ginzburg1956,GP1958}, or non-linear optics \cite{CGT1964}, among others.

The loss function is defined as the sum of two terms, which we construct according to equation \eqref{eq:loss_function} by identifying
$\mathcal{L}_v = \frac{\partial}{\partial t} - u v$, $G_v =  \frac{1}{2}\frac{\partial^2 u}{\partial x^2} + u^3$ and $\mathcal{L}_u = \frac{\partial}{\partial t} + u v$, $G_u =  - \frac{1}{2}\frac{\partial^2 v}{\partial x^2} - v^3$.

We enforce periodic boundary conditions in the x-direction and adopt identical initial conditions as those examined in  \cite{raissi2019physics}, namely $\left(u_0(x), v_0(x)\right) = \left(2 \sech (x), 0 \right)$.
Note that neither $u_0$ nor its derivatives are periodic, but they decay to zero for large $\left | x \right|$.
We extend the boundaries to $x=\pm 15$ instead of $x=\pm 5$ that was used in \cite{raissi2019physics} to ensure sufficient decay, but we keep the same limits for the time domain.
Boundary/initial conditions for $u$ and $v$ are introduced via hard-enforcement using equation \eqref{eq:hard_enforcement_periodic}:
\begin{align}
    u(t,x) &= u_0(x) + t \mathcal{N}_u \left[t,\cos \left(\frac{2 \pi x}{L}\right), \sin\left(\frac{2 \pi x}{L}\right)\right], \\
    v(t,x) &= v_0(x) + t \mathcal{N}_v \left[t,\cos \left(\frac{2 \pi x}{L}\right), \sin\left(\frac{2 \pi x}{L}\right)\right],
\end{align}

\begin{figure}[H]
    \centering
     \includegraphics[width=0.999\linewidth]{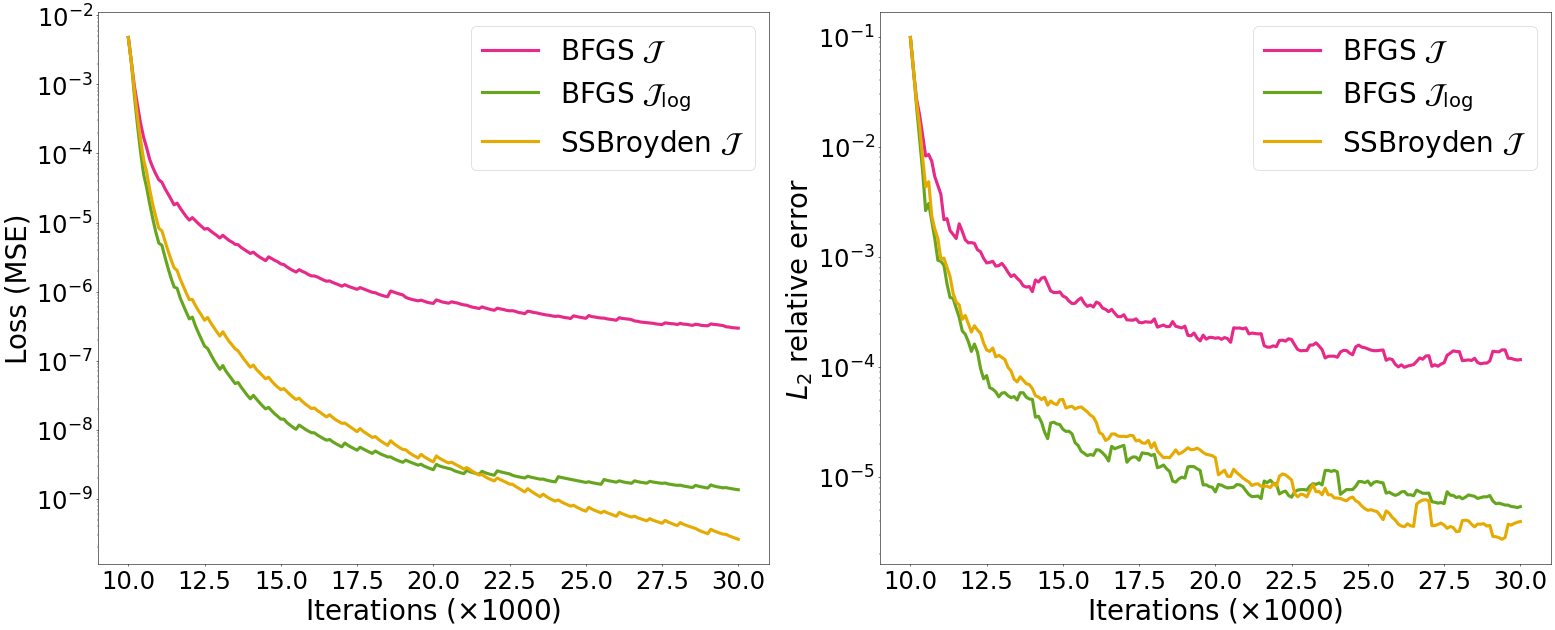}
    \caption{Convergence plots for the non-linear Schrödinger equation. Left panel: evolution of the loss function. Right panel: evolution of the relative $L_2$ error for $\left| \Psi \right|$ with respect to the reference solution on a fixed grid.}
    \label{fig:NLS_loss}
\end{figure}

\begin{figure}[H]
    \centering
    \includegraphics[width=.999\textwidth]{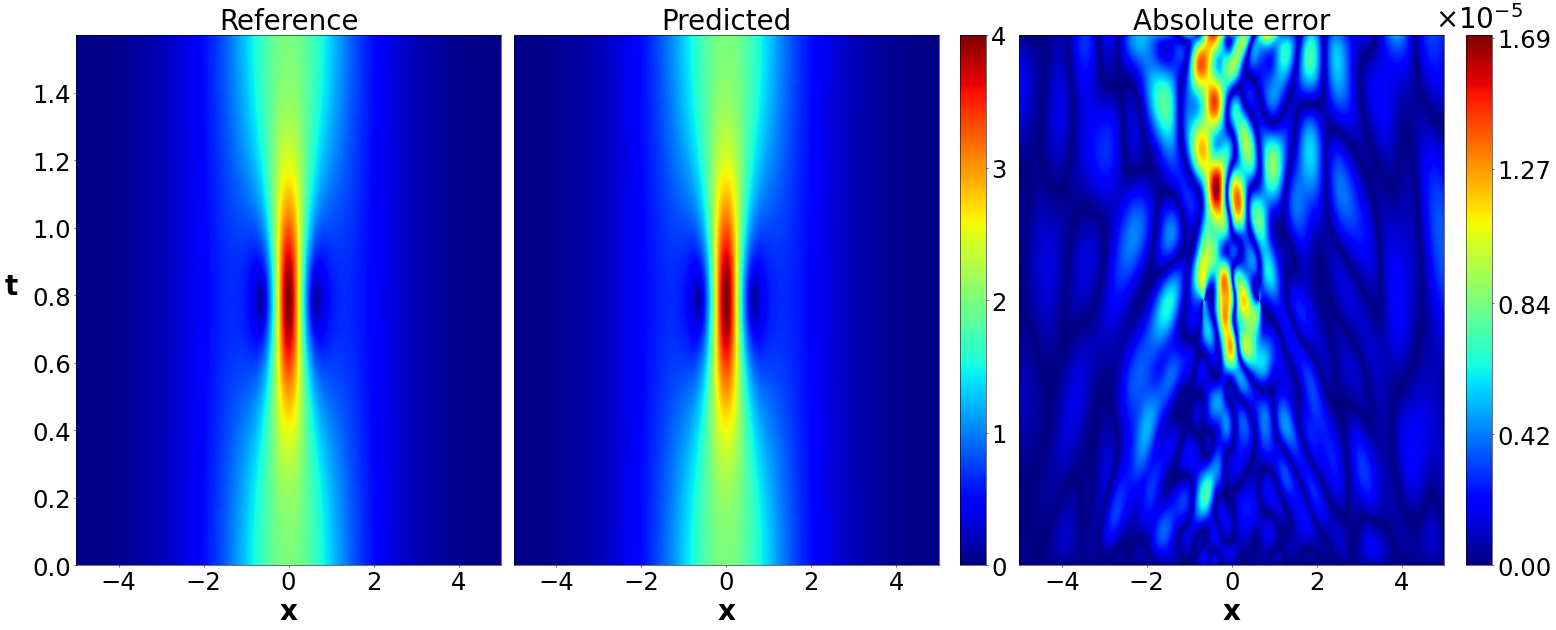} 
    \caption{Solution comparison for the non-linear Schrödinger equation. The colormaps correspond to the module of the wave function. Left panel: reference solution. Middle panel: PINN prediction obtained with the SSBroyden algorithm and the standard MSE loss. Right panel: absolute difference between the two. The solution is shown between $\left[-5,5\right]$, as outside this interval the solution decays rapidly to zero.}
    \label{fig:NLS_solution}
\end{figure}

Results for the loss function can be found in figure \ref{fig:NLS_loss}.  
In this case, while no analytical solution exists, the periodicity of the boundary conditions enables the computation of a highly accurate numerical solution using spectral methods. Specifically, we employed the Chebfun package in MATLAB (see \cite{Battles2004AnEO} for details) with a spectral Fourier discretization of 3000 modes in combination with the EDTRK4 algorithm for the temporal part (see \cite{ETDRK4}), choosing a step size of $\frac{\pi}{2}10^{-6}$. The relative $L_2$ norms between the PINN and the reference solution can be found in table \ref{tab:comparison_pinns}. Figure \ref{fig:NLS_solution} shows the numerical solution of $\left| \Psi \right|$, the PINN solution calculated employing the self-scaled Broyden algorithm, and the absolute difference between the two solutions, which are of the order of $10^{-5}$.

\paragraph{\textbf{Korteweg-De Vries equation (KdV)}}

We solve the Korteweg-De Vries (KdV) equation 
\begin{equation}\label{eq:KdV}
    \alpha \frac{\partial u}{\partial t} + \beta u \frac{\partial u}{\partial x} + \gamma \frac{\partial^3 u}{\partial x^3} = 0,
\end{equation}
where $\alpha$, $\beta$ and $\gamma$ are constants, whose standard values in the literature are $\alpha=1$, $\beta = 6$, $\gamma = 1$.

This equation characterizes the dynamics of non-linear dispersive waves, observed in shallow waters or plasma \cite{Segur1973}.
It encapsulates the fundamental principles governing these wave phenomena and has found extensive application not only in fluid mechanics but also in plasma physics \cite{Jeffrey73} and non-linear optics \cite{Horsley_2016}.
It serves as a robust test because of the presence of an important non-linear term (Burgers-like) and a third order derivative.
In this example, we reproduce a relatively difficult solution: the two-soliton solution, which can be written as 
\begin{equation}\label{eq:KdV_solution}
\small
    u_{\mathrm{an}}(x,t) = \frac{2 \left(c_1 - c_2 \right) \left[c_1 \ch^2 \left(\sqrt{c_2}\frac{\zeta_2}{2} \right) +  c_2 \sh^2 \left(\sqrt{c_1}\frac{\zeta_1}{2} \right)\right]}{\left[ \left(\sqrt{c_1} - \sqrt{c_2} \right)\ch \left( \frac{\sqrt{c_1}\zeta_1 + \sqrt{c_2}\zeta_2}{2} \right) +  \left(\sqrt{c_1} +\sqrt{c_2} \right)\ch \left( \frac{\sqrt{c_1}\zeta_1 - \sqrt{c_2}\zeta_2}{2} \right) \right]^2}.
\end{equation}
Here we denote the hyperbolic sine and cosine as $\sh$ and $\ch$ respectively and we define $\zeta_i \equiv x - c_i t - x_i$, being $c_i$ and $x_i$ arbitrary constants which describe the speed and the initial position of the solitons.
Initial and boundary conditions are given following \cite{JCM-39-816}
\begin{align}
    u(0, x) &= u_0(x), \label{eq:IC} \\
    u(t, x_0) &= g_1(t), \label{eq:BC1} \\
    u(t, x_0 + L)& = g_2(t), \label{eq:BC2} \\
    \partial_x u(t,x_0+L) &= g_3(t), \label{eq:BC3}
\end{align}
where $L$ is the size of the spacial domain and $u_0(x), g_1(t), g_2(t), g_3(t)$ are suitable functions selected to produce the analytical solution \eqref{eq:KdV_solution}.
We construct the loss function using $\mathcal{L} = \frac{\partial}{\partial t} + 6 u \frac{\partial}{\partial x} + \frac{\partial^3 }{\partial x^3}$ and $G=0$ in equation \eqref{eq:loss_function}.
We hard-enforce the Dirichlet boundary conditions by prescribing 
\begin{align}
    f_b(t,x) &= u_0(x) + A(t,x), \\
    A(t,x) &= \frac{1}{L}\left[ \left(x-x_0 \right) \left(g_2(t) - g_2(0) \right) + \left(x_0 + L - x \right) \left(g_1(t) - g_1(0) \right)\right], \\
    h_b(t,x) &= t(x-x_0)(x-x_0-L).
\end{align}
The Neumann condition \eqref{eq:BC3} is imposed via soft-enforcement as an additional term in the loss function.
The total loss function is therefore calculated as $\mathcal{J} = \mathcal{J}_{\mathrm{PDE}} + \frac{\lambda}{N_b} \sum_{i=1}^{N_b} | \frac{\partial}{\partial_x}u(t, x_0  +L) - g_3(t) |^2$, where $N_b$ is the number of points considered at the boundary and $\lambda$ is a hyperparameter to balance both terms. 
We set $N_b = 1000$ and $\lambda = 5$, as we found accurate results with these particular choices.

We choose an initial condition such that the two solitons have initial positions $x_1 = -2, x_2 = 2$ and initial velocities $c_1 = 6, c_2 = 2$.
Since  $x_1 < x_2$ but $c_1 > c_2$, eventually, the solitons will collide, triggering a non-linear interaction between them. We selected the aforementioned values to ensure that this interaction is significant and observable within the time domain under consideration.
Before or after this interaction, the solitons will travel as a linear superposition of waves (that is, as single solitons) with their respective velocities.

\begin{figure}[H]
    \centering
     \includegraphics[width=0.999\linewidth]{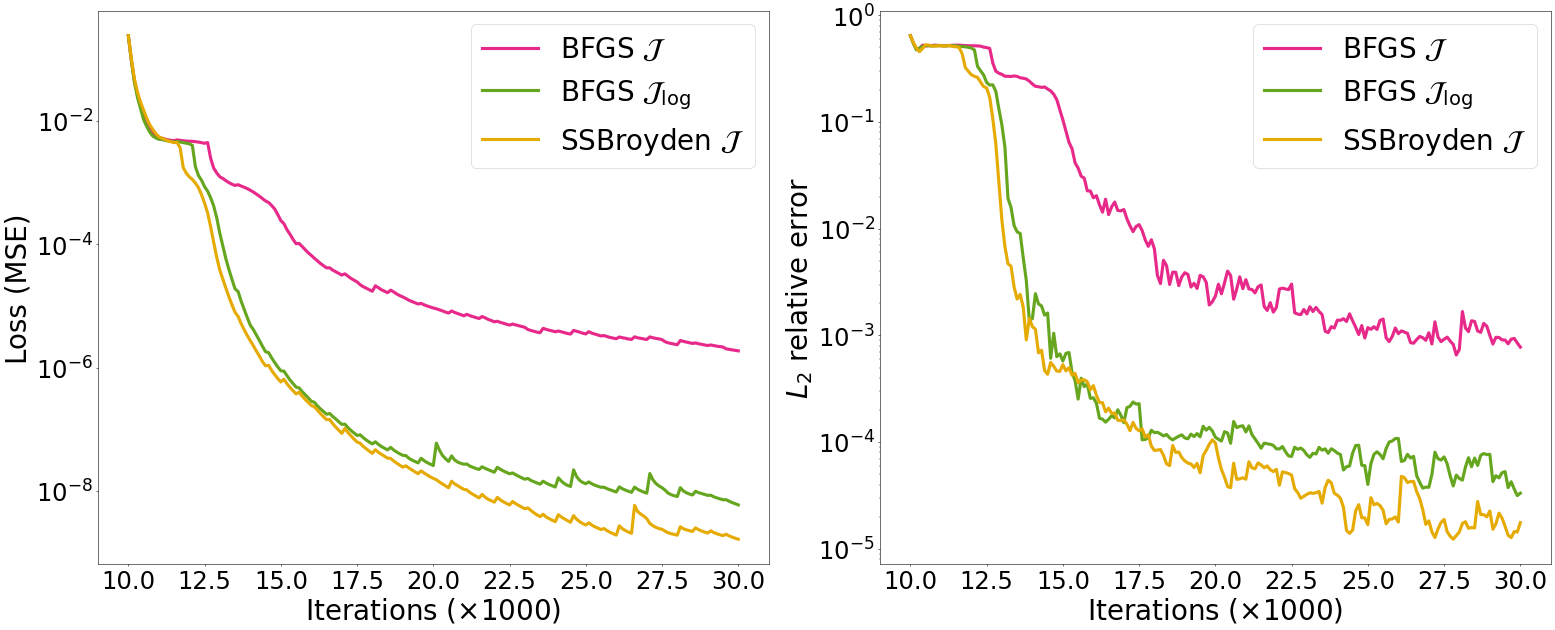}
    \caption{Convergence plots for the Korteweg-De Vries equation. Left panel: evolution of the loss function. Right panel: evolution of the relative $L_2$ error with respect to the reference solution on a fixed grid.}
    \label{fig:loss_KdV}
\end{figure}

\begin{figure}[H]
    \centering
    \includegraphics[width=.999\textwidth]{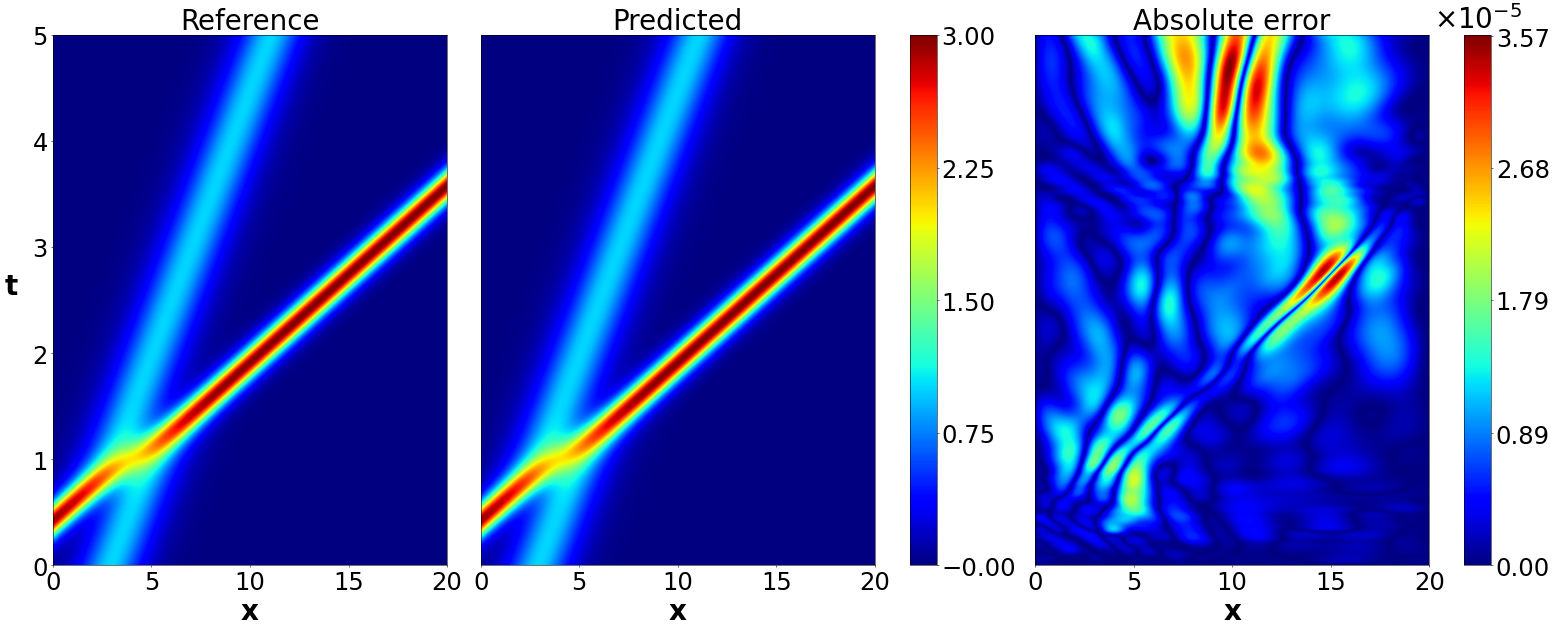} 
    \caption{Solution comparison for the Korteweg-De Vries equation. Left panel: reference solution. Middle panel: PINN prediction obtained with the SSBroyden algorithm and the standard MSE loss. Right panel: absolute difference between the two.}
    \label{fig:KdV_solution}
\end{figure}

Results for the loss function and the error norms are shown in figure \ref{fig:loss_KdV} and table \ref{tab:comparison_pinns}. Figure \ref{fig:KdV_solution} shows the reference solution, the PINN solution obtained with the self-scaled Broyden algorithm, and the absolute difference between them.

\paragraph{\textbf{One-dimensional Burgers equation (1DB)}}
A classical benchmark in the literature of PINNs (\cite{raissi2019physics,JAGTAP2020109136,ASTSK2024,HJKK2023}) is the viscous Burgers equation 
\begin{equation}\label{eq:Burgers}
    \frac{\partial u}{\partial t} + u \frac{\partial u}{\partial x} = \nu \frac{\partial^2 u}{\partial x^2},
\end{equation}

with initial data $u_0(x) = -\sin{(\pi x)}$, homogeneous boundary conditions in the domain $\left(t,x\right) \in \left[0,1\right] \times \left[-1,1 \right]$, and viscosity $\nu = 0.01/\pi$. 
Here, the input of the network consists of the coordinates $(t,x)$ in the domain, whereas the initial and the boundary conditions are imposed by setting
\begin{align}
    f_b(t,x) &= u_0(x) = - \sin \pi x, \\
    h_b(t,x) &= t(x^2 -1).
\end{align}
The loss function is given by \eqref{eq:loss_function}, where $\mathcal{L} = \frac{\partial}{\partial t} + u \frac{\partial}{\partial x} - \frac{\partial^2}{\partial x^2}$ and $G=0$. The specific set of hyperparameters for this problem can be found in table \ref{tab:training_params2}. The training set follows a random uniform distribution and is resampled every 500 iterations.

\begin{figure}[H]
    \centering
    \includegraphics[width=.999\textwidth]{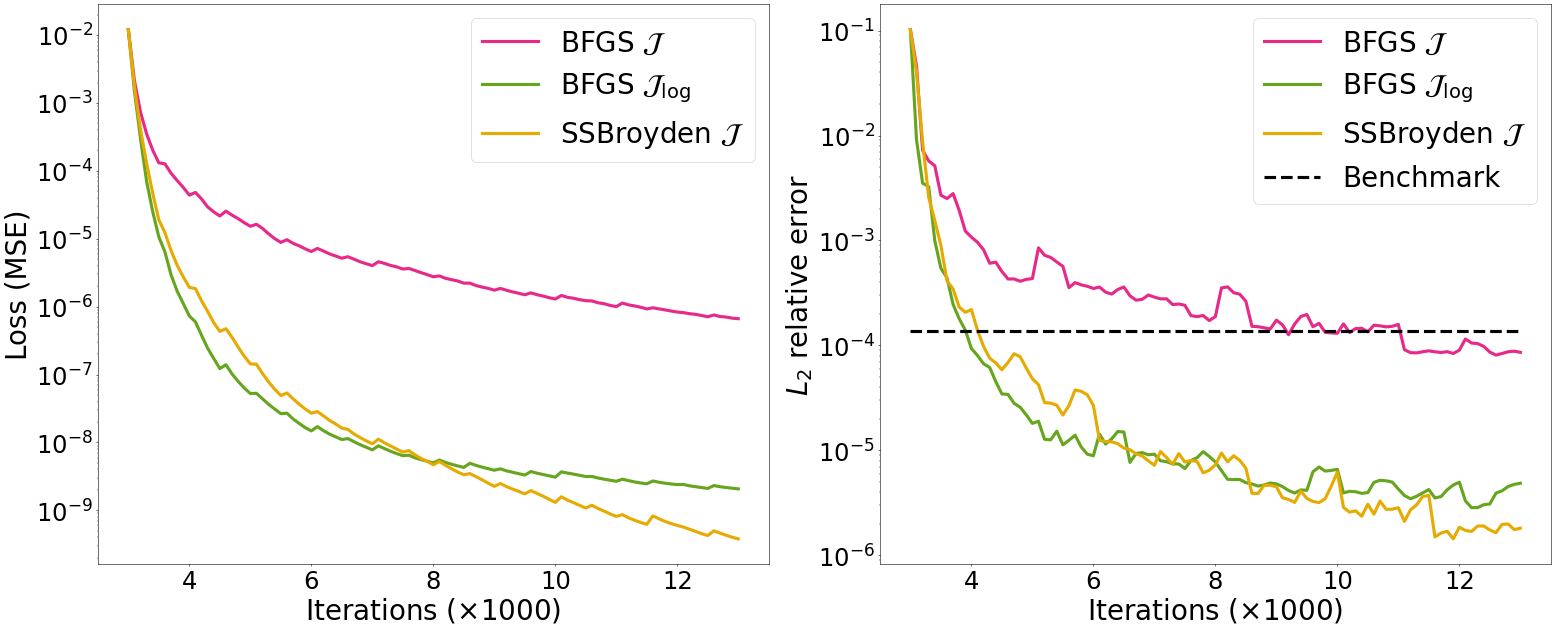} 
    \caption{Convergence plots for the Burgers equation. Left panel: evolution of the loss function. Right panel: evolution of the relative $L_2$ error with respect to the reference solution on a fixed grid. The benchmark corresponds to results reported in \cite{Wenqian2024}.}
    \label{fig:loss_burgers}
\end{figure}

\begin{figure}[H]
    \centering
    \includegraphics[width=.999\textwidth]{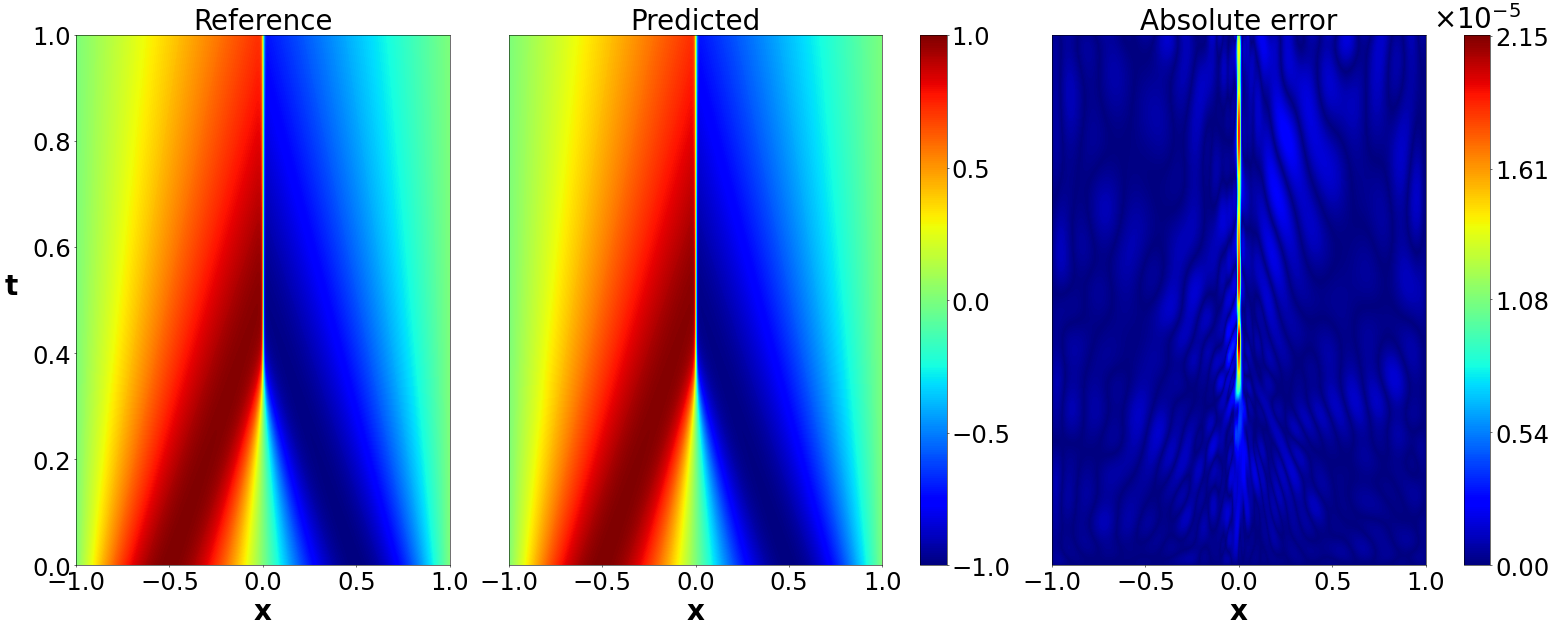} 
    \caption{Solution comparison for the Burgers equation. Left panel: reference solution. Middle panel: PINN prediction obtained with the SSBroyden algorithm and the standard MSE loss. Right panel: absolute difference between the two.}
    \label{fig:burgers_solution}
\end{figure}

Figure \ref{fig:loss_burgers} shows the evolution of the loss function and the relative $L_2$ error (left and right panels, respectively) for the three combinations of optimizer/loss function employed. The dashed horizontal line represents, to the best of our knowledge, the best result reported in the literature for this problem \cite{Wenqian2024}. 
As in other recent studies, we compare the PINN solution against a numerical solution obtained using a spectral method. 
We once again utilized the Chebfun package, with a spectral 
Fourier discretization of 2000 modes combined with the EDTRK4 algorithm for the temporal part, choosing a step size of $10^{-5}$. Final values for the relative $L_2$ errors are given in Table \ref{tab:comparison_pinns}.
Finally, figure \ref{fig:burgers_solution} shows the reference and the PINN solutions, and the absolute difference between them, employing the self-scaled Broyden algorithm together with the MSE loss. The differences are at most of the order of $\sim 10^{-5}$ and particularly concentrated where the solution is very steep.

\paragraph{\textbf{Allen-Cahn equation (AC)}} Another important benchmark in PINN literature is the Allen-Cahn equation, which is a non-linear reaction-diffusion equation that is employed to describe processes related to phase separation in binary or multi-component alloys. It is given by the following PDE:
\begin{equation}\label{eq:AC}
    \frac{\partial u}{\partial t} - \epsilon \frac{\partial^2 u}{\partial x^2} = -f(u),
\end{equation}
where $f(u)$ is a non-linear source and $\epsilon$ is constant. A popular candidate for this function is $f(u) = \kappa (u^3 - u)$, where $\kappa$ is a constant. The higher the ratio $\kappa / \epsilon$, the more pronounced the nonlinear behavior of the solution, involving sharp transitions that make the problem difficult to solve numerically. 
We consider the values $\kappa = 5$ and $\epsilon = 10^{-4}$ for the constants and the initial condition $u_0(x) = x^2 \cos \pi x$, $x \in [-1,1]$, as these are standard choices in many studies (see \cite{raissi2019physics,Wang2024PirateNetsPD,ANAGNOSTOPOULOS2024116805,CiCP-29-930,Wenqian2024}). For benchmark purposes, the initial condition is soft-enforced, whereas the periodic boundary conditions are hard-enforced by considering the Fourier functions $\left \lbrace \cos \frac{2 \pi x}{L}, \sin \frac{2 \pi x}{L} \right \rbrace$ with $L=2$, as inputs for the spatial part. It is worth noting that in some of the previously mentioned studies, the input layer includes significantly more Fourier modes than in the case presented here (see for example \cite{ANAGNOSTOPOULOS2024116805} or \cite{Wenqian2024}). The total loss function is given by
\begin{equation}\label{eq:loss_AC_cosine}
     \mathcal{J} = \mathcal{J}_{\mathrm{PDE}} + \frac{\lambda}{N_b} \lVert u - u_0 \rVert_{(t=0,x)}^2,
 \end{equation}
with $\lambda=100$. The training process consists of $5000$ Adam iterations, followed by quasi-Newton training process with improved optimizers. The rest of hyperparameters are given in table \ref{tab:training_params2}.

\begin{figure}[H]
    \centering
    \includegraphics[width=.999\textwidth]{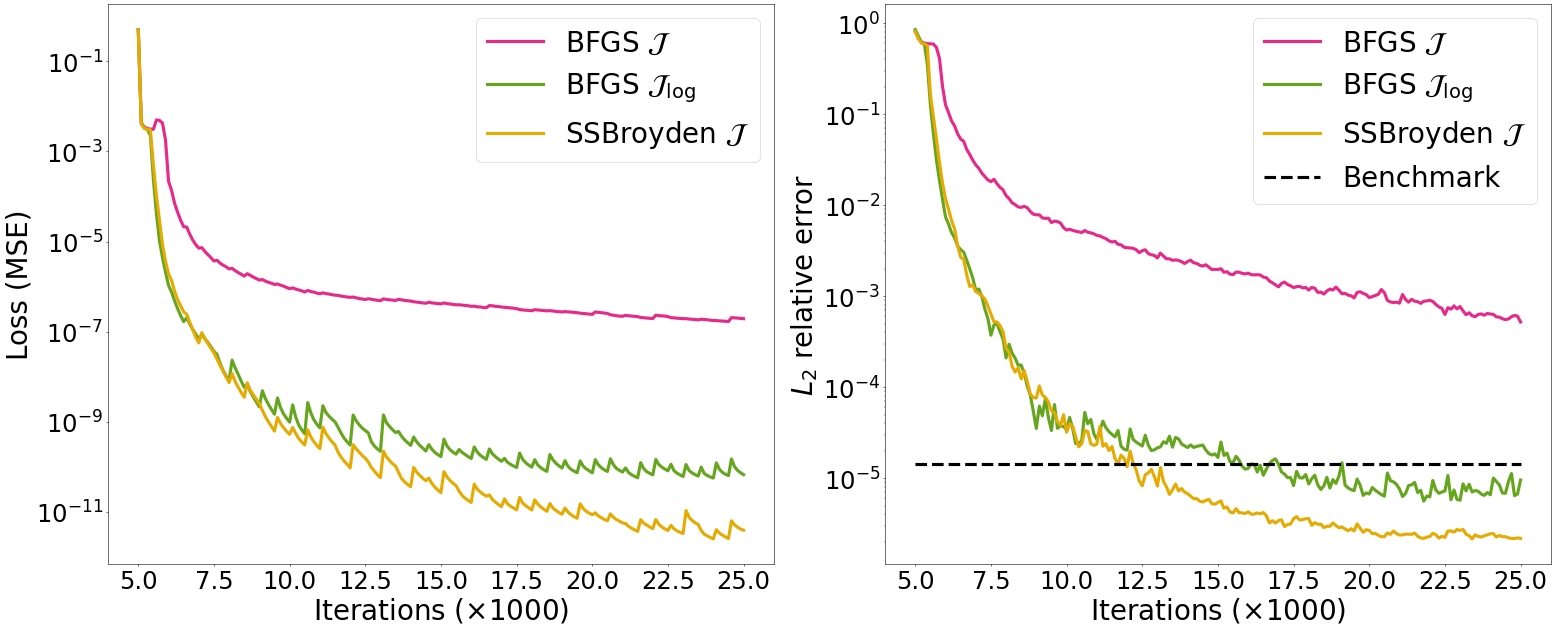} 
    \caption{Convergence plots for the Allen-Cahn equation. Left panel: evolution of the loss function. Right panel: evolution of the relative $L_2$ error with respect to the reference solution on a fixed grid. The benchmark corresponds to results reported in \cite{Wenqian2024}.}
    \label{fig:AC_cosine_loss}
\end{figure}

\begin{figure}[H]
    \centering
    \includegraphics[width=.999\textwidth]{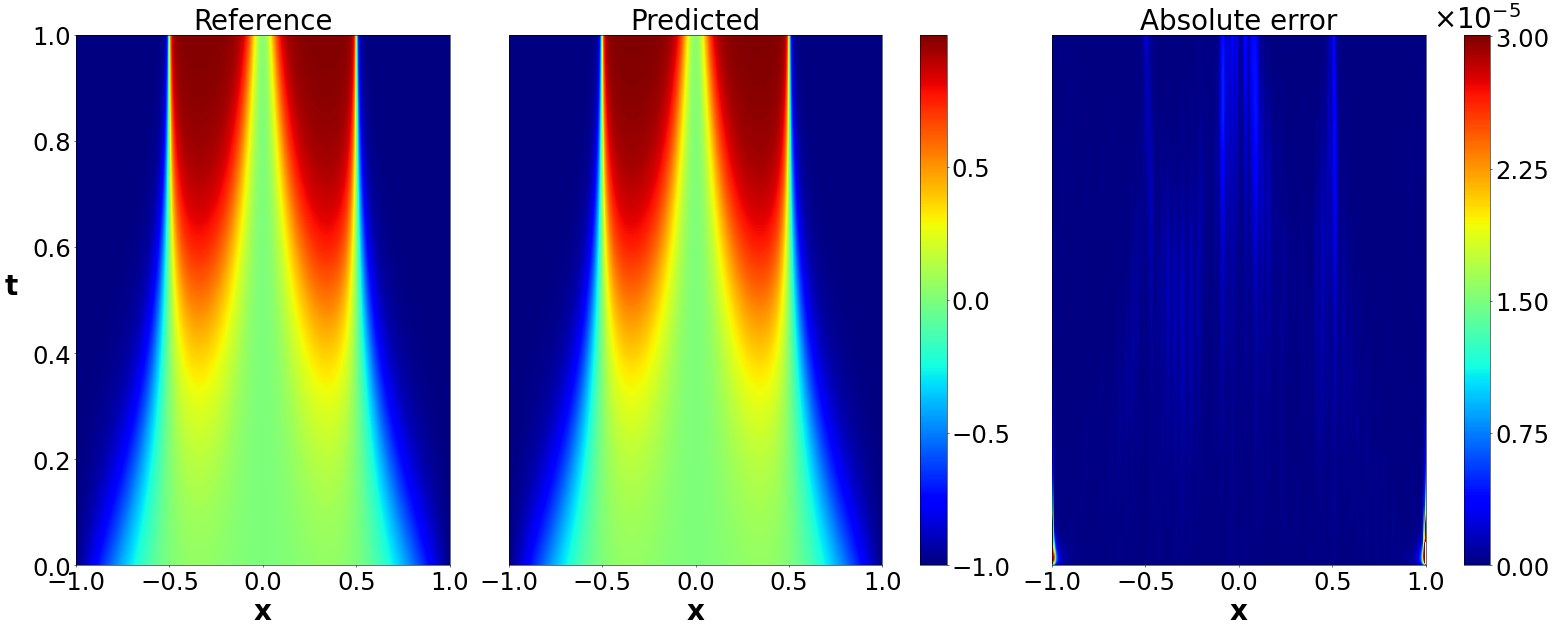} 
    \caption{Solution comparison for the Allen-Cahn equation. Left panel: reference solution. Middle panel: PINN prediction obtained with the SSBroyden algorithm and the standard MSE loss. Right panel: absolute difference between the two.}
    \label{fig:AC_solution_cosine}
\end{figure}

Figure \ref{fig:AC_cosine_loss} shows the evolution of the loss function and the $L_2$ relative error between the PINN solution and a reference numerical solution obtained with the Chebfun package with $512$ Fourier modes and the EDTRK4 algorithm with a step size of $10^{-5}$ for the temporal part (same as the one employed in \cite{ANAGNOSTOPOULOS2024116805, Wenqian2024}). The self-scaled Broyden algorithm achieves a $L_2$ relative error of $2.2 \times 10^{-6}$, which is approximately one order of magnitude lower than the best result that we found in the literature \cite{Wenqian2024}. Figure \ref{fig:AC_solution_cosine} shows colormaps of the reference solution, the PINN solution, and their absolute difference.

\paragraph{\textbf{3D Navier-Stokes: Beltrami flow (3DNS)}}

The Beltrami flow is a particular case in fluid mechanics where the vorticity vector $\bm{w} = \nabla \times \bm{u}$ is parallel to the velocity vector $\bm{u}$.
This flow satisfies the Navier-Stokes equation for an incompressible fluid:
\begin{align}
    \frac{\partial \bm{u}}{\partial t} + \left (\bm{u} \cdot \nabla \right) \bm{u} &= -\frac{1}{\rho_0}\nabla p + \nu \nabla^2 \bm{u}, \label{eq:momentum_equations} \\ 
    \nabla \cdot \bm{u} &= 0, \label{eq:incompressibility}
\end{align}
where $p$ is the pressure, $\rho_0$ is the (constant) density, and $\nu$ is the kinematic viscosity.
This is a system of four equations for four variables: the three components of the velocity vector $\bm{u} = \left(u,v,w\right)$ and the pressure $p$.
We solve it in a four dimensional domain in Cartesian coordinates $x_\alpha = (t, x, y, z)$.
A well-established, non-trivial benchmark for this problem is provided in \cite{1994IJNMF..19..369E}.
We refer the interested reader to that paper to see an illustration of this solution.
The three components of the velocity can be written in this case as
\begin{align}
    u&= -a \left [e^{ax} \sin \left(ay + dz \right) + e^{az} \cos \left(ax + dy  \right) \right]e^{-d^2 t}, \\
    v&= -a \left [e^{ay} \sin \left(az + dx \right) + e^{az} \cos \left(ay + dz  \right) \right]e^{-d^2 t}, \\
    w&= -a \left [e^{az} \sin \left(ax + dy \right) + e^{ay} \cos \left(az + dx  \right) \right]e^{-d^2 t}.
\end{align}
for arbitrary constants $a, d$. 
The solution for the pressure can be written as
\begin{equation}
    \begin{split}
        p = -\frac{a^2}{2} \big[ & e^{2ax} + e^{2ay} + e^{2az} + 2 \sin{(ax + dy)} \cos{(az + dx)} e^{a(y+z)} \\ 
        &+ 2 \sin{(ay + dz)} \cos{(ax + dy)} e^{a(z+x)} \\
        &+ 2 \sin{(az + dx)} \cos{(ay + dz)} e^{a(y+x)} \big] e^{-2d^2 t}.
    \end{split}\label{eq:analyticalp}
\end{equation}
In the usual way, we can construct the loss function as the sum of four terms.
For each component of equation \eqref{eq:momentum_equations} we have $\mathcal{L}_i = \frac{\partial}{\partial t} + \left(\bm{u} \cdot \nabla \right) - \nu \nabla^2$ and $G = - \frac{1}{\rho_0}\nabla p$, whereas for equation \eqref{eq:incompressibility} $\mathcal{L} = \mathrm{div}$ and $G=0$.

We impose Dirichlet boundary conditions for the three components of the velocity vector $\bm{u}$.
We will describe them for one of them, without loss of generality.
If the spatial domain is $\left[x_0, x_0 + L_x \right] \times \left[y_0, y_0 + L_y \right] \times \left[z_0, z_0 + L_z \right]$, we define
\begin{align*}
    f_0(y,z,t) &= u(x_0,y,z,t) - u_0(x_0,y,z), \\
    f_1(y,z,t) &= u(x_0 + L_x,y,z,t) - u_0(x_0+L_x,y,z), \\
    g_0(x,z,t) &= u(x,y_0,z,t) - u_0(x,y_0,z), \\
    g_1(x,z,t) &= u(x,y_0 + L_y,z,t) - u_0(x,y_0 + L_y,z), \\
    h_0(x,y,t) &= u(x,y,z_0,t) - u_0(x,y,z_0), \\
    h_1(x,y,t) &= u(x,y,z_0+L_z,t) - u_0(x,y,z_0+L_z)
\end{align*}
Then, the functions $f_b$ and $h_b$ used in \eqref{eq:hard_enforcement_dirichlet} can be defined through the following ansatz
\begin{align}
    f_b(x,y,z,t) &= u_0(x,y,z) + A(x,y,z,t), \label{eq:fb3D} \\
    h_b(x,y,z,t) &= t\prod_{w = x,y,z}\left(w-w_0 \right) \left(w-w_0-L_w \right), \label{eq:h3D} \\
    \begin{split}
        A(x,y,z,t) &= \left(1-\xi_x\right) f_0(y,z,t) + \xi_x f_1(y,z,t) + \left(1-\xi_y \right)G_0(x,z,t) \\
        &+ \xi_y G_1(x,z,t) + \left(1-\xi_z \right)H_0(x,y,t) + \xi_z H_1(x,y,t),
    \end{split}\label{eq:A3D}
\end{align}
where $\xi_q = \frac{q - q_0}{L_q}$ ($q = x,y,z$) and the functions $\left \lbrace G_i, H_i \right \rbrace_{i=0,1}$ are defined as
\begin{align}
    &G_i(x,z,t) = g_i(x,t) - \left(1-\xi_x \right) g_i(x_0,t) - \xi_x g_i(x_0 + L_x,t), \\ 
    \begin{split}
        &H_i(x,y,t) = h_i(x,y,t) - \left(1-\xi_x \right)h_i(x_0,y,t) - \xi_x h_i(x_0 + L_x,y,t) \\
        &- \left(1-\xi_y \right) \left \lbrace h_i(x,y_0,t) - \left(1-\xi_x \right)h_i(x_0,y_0,t) - \xi_x h_i(x_0 + L_x,y_0,t)\right \rbrace \\
        &- \xi_y \left \lbrace h_i(x,y_0 + L_y,t) - \left(1-\xi_x \right)h_i(x_0,y_0 + L_y,t) - \xi_x h_i(x_0 + L_x,y_0 + L_y,t)\right \rbrace. \\
    \end{split}
\end{align}

\begin{figure}[tbp!]
    \centering
    \includegraphics[width=0.999\textwidth]{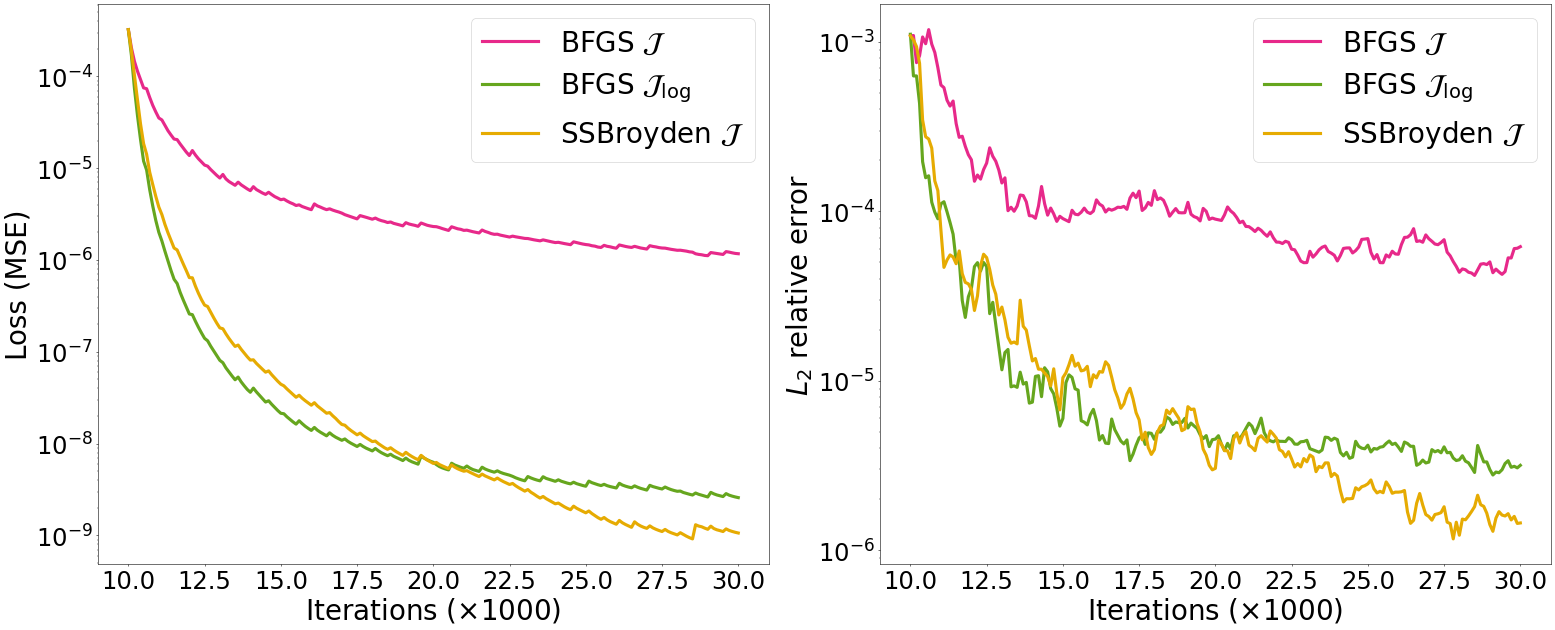}
    \caption{Convergence plots for the 3D Navier-Stokes equation. Left panel: evolution of the loss function. Right panel: evolution of the relative $L_2$ error with respect to the reference solution on a fixed grid.}
    \label{fig:loss_Beltrami}
\end{figure}

\begin{figure}[tbp!]
    \includegraphics[width=0.999\textwidth]{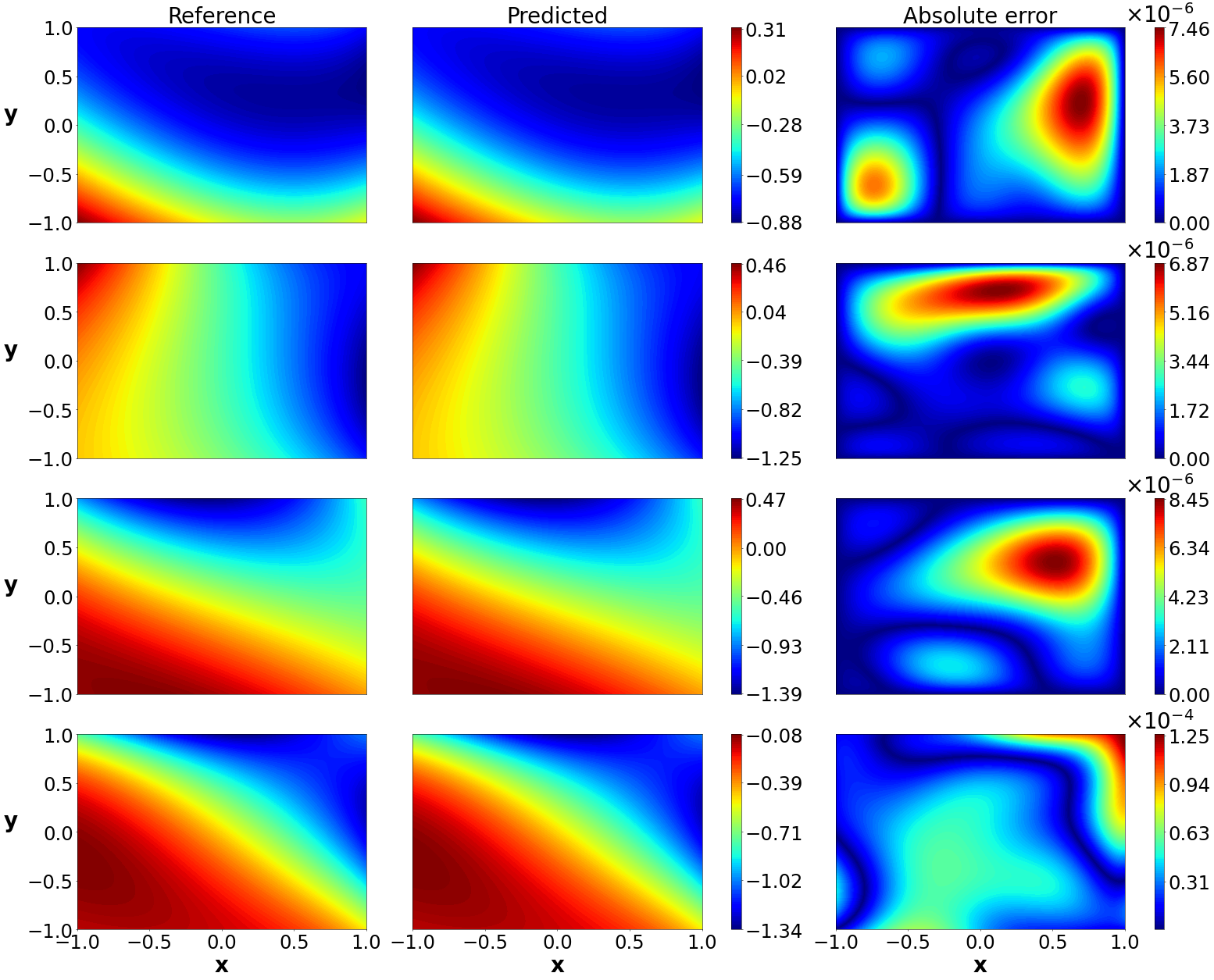} 
    \caption{Solution comparison for the 3D Navier-Stokes equation. Left panels: reference solution. Middle panels: PINN prediction obtained with the SSBroyden algorithm and the standard MSE loss. Right panels: absolute difference between the two. The plots show cuts of the $x-y$ plane at $z = 0.5$ and $t = 1$. The first three rows show the velocity components. The fourth row shows the pressure.}
    \label{fig:Beltrami_solution}
\end{figure}

Regarding the pressure, we only need to specify it at a single spacial point $(x_1, y_1, z_1)$ for all times $t$.
This is necessary because one has the freedom to add a function of time to $p$ and get an equivalent solution.
Indeed, if we replace $p \rightarrow p + f(t)$ in equation \eqref{eq:momentum_equations} we obtain the same system and the solution is ambiguous. 

Figure \ref{fig:loss_Beltrami} and table \ref{tab:comparison_pinns} show the results for the loss function and the error norms. Figure \ref{fig:Beltrami_solution} shows colormaps in a given cut of the components of the velocity and the pressure predicted with the PINN using the self-scaled Broyden algorithm, together with the analytical ones and the absolute difference between them.

\paragraph{\textbf{Lid-driven cavity (LDC)}}

To conclude, we present the results obtained for the 2D lid-driven cavity flow treated in recent studies \cite{Wang2024PirateNetsPD, WTP21}. This problem is governed by the stationary momentum equation
\begin{equation}
    \left (\bm{u} \cdot \nabla \right) \bm{u} = -\nabla p + \nu \nabla^2 \bm{u},
\end{equation}
together with the continuity equation for an incompressible flow
\begin{equation}
    \nabla \cdot \bm{u} = 0,
\end{equation}
where $\bm{u} = \left(u(x,y),v(x,y)\right)$ is the vector velocity and $p$ denotes the pressure. The coefficient $\nu$ is commonly expressed in terms of the Reynolds number $\mathrm{Re}$ by setting $\nu = 1/\mathrm{Re}$.

The domain is the square $\left[0,1 \right]^2$. The boundary conditions for this problem are
\begin{equation}
    \bm{u}_b (x,y) =
    \begin{cases}
        &\left(u_T (x),0 \right), \quad y=1, \\
        &\left(0,0 \right), \quad \mathrm{otherwise}
    \end{cases}
\end{equation}
where $u_T(x)$ must equal $1$ across most of the top edge of the domain and decay smoothly to zero at the corners to meet the Dirichlet boundary conditions in the horizontal direction. More concretely, we consider the same function as in \cite{Wang2024PirateNetsPD}
\begin{equation}
    u_T(x) = 1 - \frac{\cosh{\left(C_0\left(x-0.5\right)\right)}}{\cosh{\left(0.5 C_0 \right)}}
\end{equation}
where $C_0=50$. Following \cite{WTP21,ASTSK2024}, we employ the stream function formalism, solving for $\psi$ instead of the two components of the velocity. The pressure is the other output of the network and is set to zero at a particular point in the domain to avoid ambiguities in the solution. The total loss function is then given in this case by
\begin{equation}
    \mathcal{J} = \mathcal{J}_{\mathrm{PDE}} + \frac{\lambda}{N_b} \lVert \bm{u} - \bm{u}_b \rVert_{(x,y) \in \mathcal{B}}^2,
\end{equation}
where $\mathcal{B}$ is the boundary of $\left[0,1 \right]^2$, and $\mathcal{J}_{\mathrm{PDE}}$ is the sum of the residuals corresponding to the two momentum equations, constructed with \eqref{eq:loss_function} by identifying $\mathcal{L}_u = \frac{\partial}{\partial t} - \nu \nabla^2 u$, $G_u = -\left(\bm{u} \cdot \nabla \right)u - \frac{\partial p}{\partial x}$ and $\mathcal{L}_v = \frac{\partial}{\partial t} - \nu \nabla^2 v$, $G_v = -\left(\bm{u} \cdot \nabla \right)v - \frac{\partial p}{\partial y}$. The following results are obtained by setting $\mathrm{Re} = 1000$. The parameter $\lambda$ that controls the contribution of the boundary conditions is set to $10$.

\begin{figure}[H]
    \centering
     \includegraphics[width=0.7\linewidth]{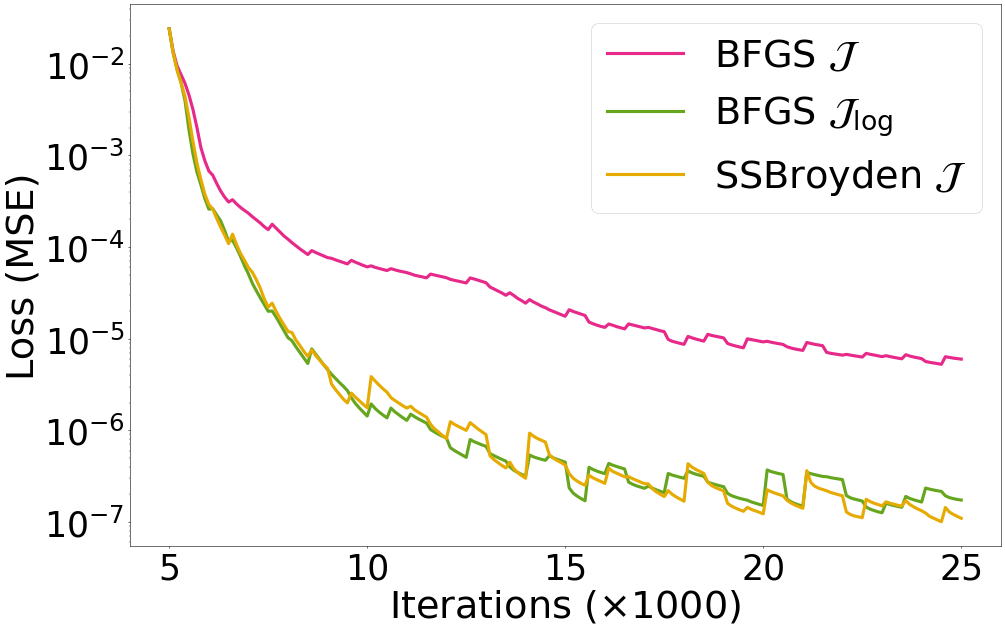}
    \caption{Loss functions for the lid-driven cavity problem.}
    \label{fig:loss_LDC}
\end{figure}

\begin{figure}[tbp!]
    \includegraphics[width=0.999\textwidth]{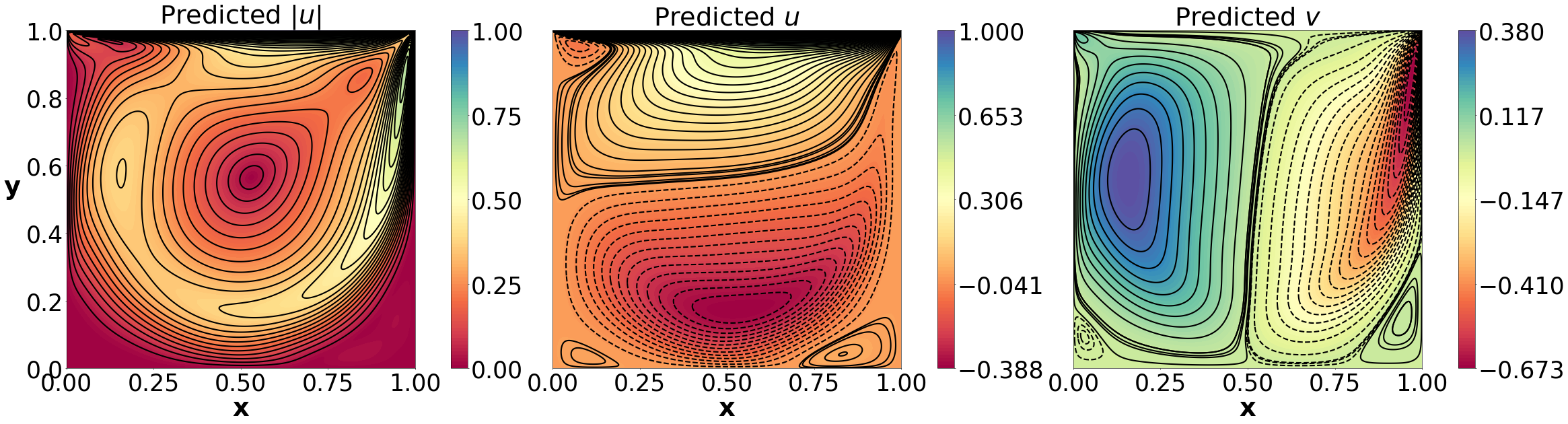} 
    \caption{PINN predictions for the lid-driven cavity problem. The plots show the components of the vector velocity $\bm{u}$ and its modulus. The lines correspond to contours of the quantities represented in each plot.}
    \label{fig:LDC_solution}
\end{figure}

Results for the loss function are shown in figure \ref{fig:loss_LDC}.
Figure \ref{fig:LDC_solution} shows the solution for the components and the modulus of the velocity acquired with the self-scaled Broyden algorithm. Since no analytical or high-precision numerical solution is available, we estimate the error according to equation \eqref{eq:eNN_definition} as
$\epsilon_{\mathrm{NN}} \approx 1.3 \times 10^{-4}$, obtained with the self-scaled Broyden algorithm. The standard BFGS algorithm on the other hand achieves, if trained against the standard MSE loss $\mathcal{J},$ a value of $\epsilon_{\mathrm{NN}} \approx 1.0 \times 10^{-3}$, whereas if the logarithm of the MSE loss $\mathcal{J}_{\mathrm{log}}$ is employed, this value is reduced up to $\epsilon_{\mathrm{NN}} \approx 1.6 \times 10^{-4}$.

\subsection{Comparison with the literature}
Wrapping up this section, in table \ref{tab:comparison_pinns} we offer a direct one-to-one comparison of our findings (marked as TW --this work--) presented earlier with similar problems encountered in the PINNs literature (specific references provided in the second column). 
We utilized the code provided in the GitHub repository cited in \cite{JCM-39-816} for the KdV equation, incorporating the analytical 2-soliton solution and defining the same training domain for this study. 
As for the NLS equation, we employed our code (as the L-BFGS optimizer used in \cite{raissi2019physics} is the same as the one considered here) to compute the corresponding PINN approximation errors defined in \ref{sssec:force_free}. Finally, for the Allen-Cahn equation we have employed the code given in the Github repository cited in \cite{Wenqian2024}. 

In the current landscape of PINN research, L-BFGS has become a widely used optimizer due to its generally superior performance compared to the Adam optimizer. However, it has demonstrated some limitations, especially when confronted with highly ill-conditioned problems.
Consequently, it needs a significantly higher number of trainable parameters than the BFGS algorithm to achieve comparable precision. 
Although training two identical networks with BFGS would be markedly slower compared to L-BFGS, primarily due to the former's requirement to store and update the inverse Hessian estimate at each iteration, with BFGS one can use much smaller networks with higher accuracy. 

\begin{figure}[h!]
    \centering
    \begin{subfigure}[t]{0.5\textwidth}
        \centering
        \includegraphics[width=\textwidth]{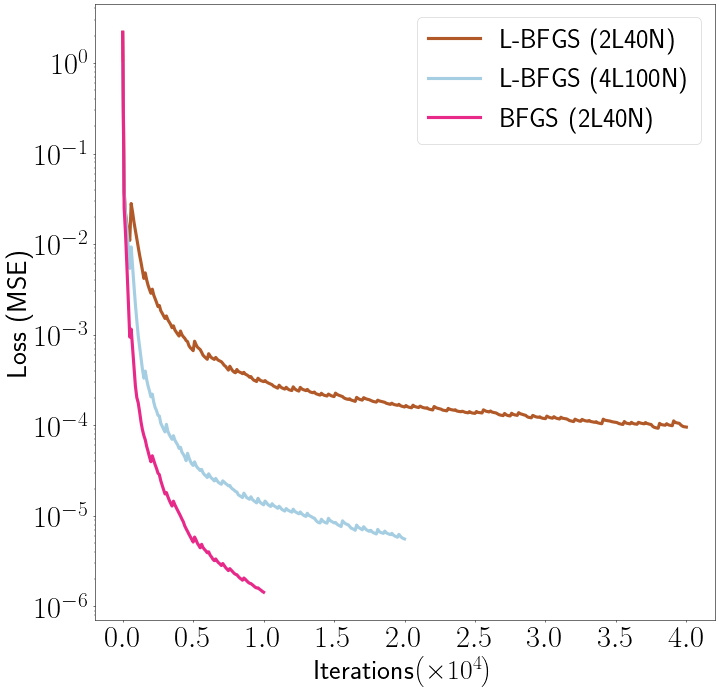}
        \caption{}\label{fig:comparison_lbfgs_bfgs_epochs}
    \end{subfigure}%
    ~
    \begin{subfigure}[t]{0.5\textwidth}
        \centering
        \includegraphics[width=\textwidth]{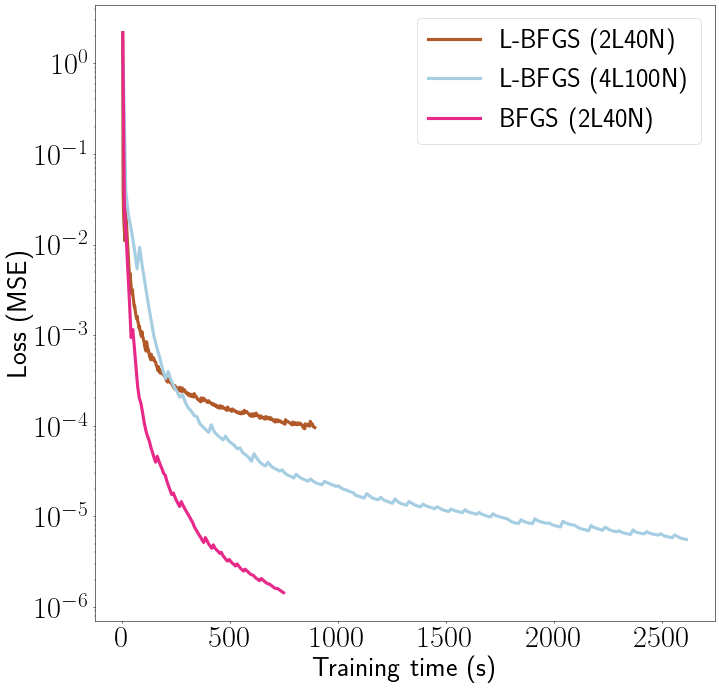}
        \caption{}\label{fig:comparison_lbfgs_bfgs_time}
    \end{subfigure}
    \caption{Loss function obtained for the non-linear Schrödinger equation considering the L-BFGS and BFGS algorithms with the network employed in this work (2L40N) and the L-BFGS algorithm for the model suggested in \cite{raissi2019physics} (4L100N), as a function of the number of iterations (\ref{fig:comparison_lbfgs_bfgs_epochs}) and the training time (\ref{fig:comparison_lbfgs_bfgs_time}).}
    \label{fig:comparison_lbfgs_bfgs}
\end{figure}

Figure \ref{fig:comparison_lbfgs_bfgs} illustrates this effect. We show the evolution of $\mathcal{J}$ as a function of the number of iterations (left panel) or training time (right panel) for the NLS equation. We compare the training process with the L-BFGS and the BFGS algorithms using our smaller neural network (see table \ref{tab:training_params}), and the L-BFGS algorithm with the same NN considered in \cite{raissi2019physics}. 

For identical networks, L-BFGS runs about four times faster, but its convergence rate is significantly slower. Using a larger network with L-BFGS improves accuracy but significantly increases training time. Overall, to achieve a given accuracy, BFGS proves more efficient than L-BFGS for the same number of parameters, both in terms of iterations and total training time.

\begin{figure}[h!]
    \centering
    \includegraphics[width=.7\textwidth]{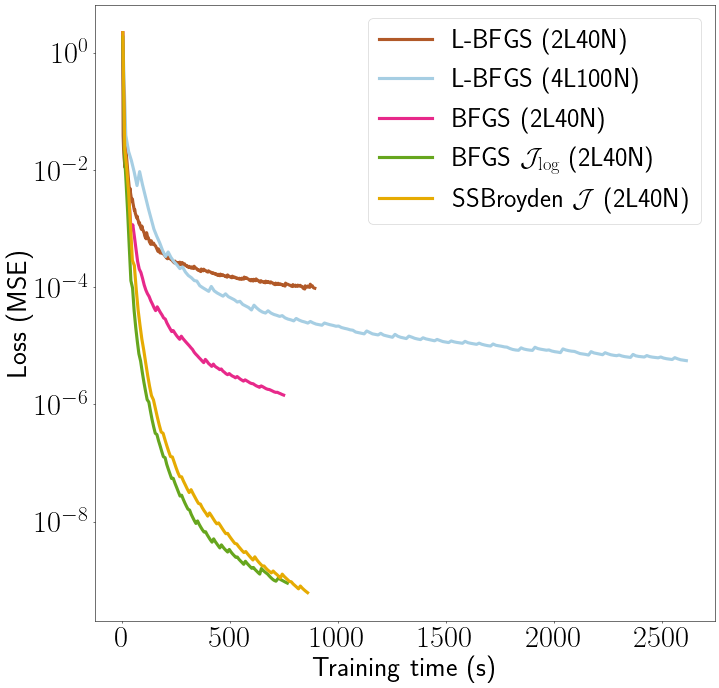}
    \caption{Loss function vs. training time for the non-linear Schrödinger equation obtained with our training process modifications. The results presented in figure \ref{fig:comparison_lbfgs_bfgs_time} are also shown for reference.}
    \label{fig:comparison_lbfgs_bfgs_modifications}
\end{figure}

However, the preceding examples in this section illustrate that training with BFGS in conjunction with the MSE loss was not the optimal strategy either. By introducing modifications to either the BFGS algorithm or the loss function, the performance enhancement becomes even more remarkable, as shown in Figure \ref{fig:comparison_lbfgs_bfgs_modifications}.

With these adjustments, we observe a reduction of several orders of magnitude in the loss function within the same training duration, compared to L-BFGS. This is directly reflected in the errors of the solution achieved in each case, summarized in table \ref{tab:comparison_pinns}.
The errors computed in this work are averaged over multiple training runs, employing different random initializations of the trainable parameters to ensure robustness. Our results are presented in the format (mean ± standard deviation), to illustrate both, the average error and the variability across the different trials. In all cases, our solutions are 1-2 orders of magnitude more precise than the references in the literature, although we are using significantly smaller networks (see columns 5-6 in the table).

\begin{table}[H]
    \centering
    \small
    \caption{Comparison between models in the literature (references in the second column) and our models in this work (TW), with modifications in the optimizer and loss function. Errors for a given variable $x$ are denoted by $E_x$ and are computed as the $L_2$ relative norm of the difference with a reference analytical or numerical solution, (see equation \eqref{eq:error_norm}). 
    }
    \begin{tabular}{c c c c c c c c c c c}
        \hline 
        Problem & Reference & Optimizer & Loss & Layers & Neurons & Error
        \\ \hline
        \multirow{5}{*}{\shortstack{2DH \\ $(1,4)$}} & & & & & & $E_u$
        \\ [0.5ex]
        & \cite{ANAGNOSTOPOULOS2024116805} & L-BFGS & $\mathcal{J}$ & 6 & 128 & $8.21 \times 10^{-6}$ \\ [0.5ex]
        & TW & SSBroyden & $\mathcal{J}$ & 2 & 20 & $(6 \pm 4) \times 10^{-8}$\\ [0.5ex]
        & TW &  BFGS & $\mathcal{J}_{\mathrm{log}}$ & $2$ & 20 & $ (3.6 \pm 0.2) \times 10^{-7}$ \\ \hline
        \multirow{5}{*}{\shortstack{NLP\\ $(k=1)$}} & & & & & & $E_{\phi}$
        \\ [0.5ex]
        & \cite{SharmaShankar2022} & L-BFGS & $\mathcal{J}$ & 4 & 50 & $1.08 \times 10^{-6}$ \\ [0.5ex]
        & TW & SSBroyden & $\mathcal{J}$ & 2 & 30 & $(3 \pm 1) \times 10^{-9}$\\ [0.5ex]
        & TW &  BFGS & $\mathcal{J}_{\mathrm{log}}$ & $2$ & 30 & $(6\pm 3) \times 10^{-9}$ \\ \hline
        \multirow{5}{*}{NLS} & & & & & & $\left( E_{u},E_{v} \right)$ \\ [1ex]
         & \cite{raissi2019physics} & L-BFGS & $\mathcal{J}$ & 4 & 100 & $(2.23, 2.05) \times 10^{-3}$ \\ [0.5ex]
        & TW & SSBroyden & $\mathcal{J}$ & 2 & 40 & $(5 \pm 1, 8 \pm 2) \times 10^{-6}$ \\ [0.5ex]
        & TW & BFGS & $\mathcal{J}_{\mathrm{log}}$ & $2$ & 40 & $(1 \pm 0.1, 1.5 \pm 0.3) \times 10^{-5}$ \\
        \hline
        \multirow{6}{*}{KdV} & & & & & & $E_{u}$ \\ [0.5ex]
        & \cite{JCM-39-816} & L-BFGS & $\mathcal{J}$ & 4 & 32 & $1.07 \times 10^{-2}$ \\ [0.5ex]
         & \cite{JCM-39-816} & L-BFGS & $\mathcal{J}$ & 8 & 60 & $1.26 \times 10^{-3}$\\ [0.5ex]
        & TW & SSBroyden & $\mathcal{J}$ & 3 & 30 & $ (6\pm 0.7) \times 10^{-6}$ \\ [0.5ex]
        & TW & BFGS & $\mathcal{J}_{\mathrm{log}}$ & $3$ & 30 & $(1.2 \pm 0.4) \times 10^{-5}$  \\
        \hline
         \multirow{5}{*}{1DB} & & & & & & $E_{u}$ \\ [0.5ex]
        & \cite{Wenqian2024} & Adam & $\mathcal{J}$ & 6 & 128 & $4.8 \times 10^{-4}$ \\ [0.5ex]
        & TW & SSBroyden & $\mathcal{J}$ & 3 & 20 & $(2.9 \pm 0.4) \times 10^{-6}$ \\ [0.5ex]
        & TW & BFGS & $\mathcal{J}_{\mathrm{log}}$ & $3$ & 20 & $(5 \pm 2) \times 10^{-6}$  \\
        \hline
        \multirow{5}{*}{AC} & & & & & & $E_{u}$ \\ [0.5ex]
        & \cite{Wenqian2024} & Adam & $\mathcal{J}$ & 6 & 128 & $1.45 \times 10^{-5}$ \\ [0.5ex]
        & TW & SSBroyden & $\mathcal{J}$ & 3 & 30 & $ (2.2 \pm 0.7) \times 10^{-6}$ \\ [0.5ex]
        & TW & BFGS & $\mathcal{J}_{\mathrm{log}}$ & $3$ & 30 & $(9.7 \pm 0.8) \times 10^{-6}$  \\
        \hline
        \multirow{6}{*}{3DNS} & & & & & & $\left(E_{u},E_{v},E_{w} \right)$ \\ [0.5ex]
         & \cite{JIN2021109951} & L-BFGS & $\mathcal{J}$ & 7 & 50 & $(2.54, 2.40, 2.60) \times 10^{-5}$ \\ [0.5ex]
         & \cite{Wang2024} & L-BFGS & $\mathcal{J}$ & 5 & 50 & $(2.64, 4.35, 2.74) \times 10^{-5}$ \\ [0.5ex]
        & TW & SSBroyden & $\mathcal{J}$ & 2 & 40 & $(7.3 \pm 0.9, 7.1 \pm 0.4, 7.8 \pm 0.1) \times 10^{-7}$\\ [0.5ex]
        & TW & BFGS & $\mathcal{J}_{\mathrm{log}}$ & $2$ & 40 & $(1.3 \pm 0.1, 1.4 \pm 0.1, 1.3 \pm 0.1) \times 10^{-6}$  \\
        \hline
    \end{tabular}
    \label{tab:comparison_pinns}
\end{table}

\section{Conclusions.}\label{sec:conclusions}

In this study, we have investigated the boundaries of accuracy achievable by physics-informed neural networks (PINNs). Unlike conventional methods, which benefit from a solid mathematical foundation (built over many decades) offering insights into methodological order and accuracy constraints, the young field of PINNs usually relies on a brute force methodology involving trial and error. We emphasize the pivotal significance of the optimization algorithm in attaining robust convergence, irrespective of the specific physical problem. Furthermore, we demonstrate how making appropriate selections can substantially enhance result accuracy by several orders of magnitude, independently of the specific physical problem being addressed.

In the family of quasi-Newton methods, the convergence rate of each algorithm is linked to the well-conditioning of the corresponding Hessian matrix $\mathrm{hess}(\mathcal{J}(\bm{z}))$.
We have demonstrated that,
when the eigenvalue spectrum of $\mathrm{hess}(\mathcal{J}(\bm{z}))$ is centered around unity with minimal dispersion, the optimization process efficiently minimizes the loss function, resulting in highly precise solutions.
A similar effect is produced by considering a modified loss function $\mathcal{J}_g$ instead of the usual MSE loss.
Changes to the loss function can be paired with selecting an optimization algorithm, enabling thorough exploration of different combinations. Our research suggests that the optimization algorithm choice typically has a greater effect on convergence compared to adjustments to the loss function. However, the ease of implementation may favor employing modifications to the loss function for practical purposes: very often it is much easier to tweak the loss function than to either change an existing optimizer or create a new one entirely.

However, as the field of PINNs continues to evolve and we gain a deeper understanding of the optimization process underlying neural network training, more sophisticated algorithms will likely become readily available in popular machine learning frameworks. This study also seeks to encourage developers to move in this direction.

A crucial result of refining the optimization process in PINNs is the significant reduction in both the size and complexity of the networks used to tackle similar problems. Through our series of benchmarks, we have demonstrated how various problems from the literature can be solved with notably smaller network sizes and improved precision, as illustrated in Table \ref{tab:comparison_pinns}. This enhances numerical efficiency, addressing a key weakness of PINNs compared to classical numerical methods.

Indeed, all simulations presented in this study were conducted on a standard PC or, in some cases, on a regular laptop, without any specialized requirements. While enhancements to other hyperparameters such as activation functions and network structure can provide further assistance, our primary finding underscores the crucial importance of leveraging improved optimizers and rescaled loss functions.
We anticipate that this effect will become even more significant as the problem's dimensionality increases, especially when addressing large-scale problems. 

Currently, the most frequently employed optimizer in PINNs is the L-BFGS algorithm, which is a faster variant of BFGS (per iteration) but suffers even more from ill-conditioning.
As a consequence, the latter needs much more trainable parameters to obtain results of similar accuracy.
As future work, we will explore how the different modifications of the L-BFGS algorithm suggested in optimization theory literature affect the convergence in PINNs.

\section*{Acknowledgments}

We acknowledge the support through the grant PID2021-127495NB-I00 funded by MCIN/AEI/10.13039/501100011033 and by the European Union, the Astrophysics and High Energy Physics programme of the Generalitat Valenciana ASFAE/2022/026 funded by MCIN and the European Union NextGenerationEU (PRTR-C17.I1), and the Prometeo excellence programme grant CIPROM/2022/13. JFU is supported by the predoctoral fellowship ACIF 2023, cofunded by Generalitat Valenciana and the European Union through the European Social Fund.

\appendix

\section{Derivation of the Grad-Shafranov equation}\label{app:gs_derivation}

In axisymmetry, the magnetic field $\bm{B}$ can be described in terms of two poloidal and toroidal stream functions $\mathcal{P}$ and $\mathcal{T}$ as
\begin{equation}\label{eq:poloidal_toroidal}
     \bm{B} = \frac{q}{\sqrt{1-\mu^2}} \left( \nabla\mathcal{P} \times \bm{e}_\phi + \mathcal{T} \bm{e}_\phi \right).
\end{equation}
where $\bm{e}_\phi$ is the unit vector in the $\phi$ direction.

Substituting this expression into the force-free condition $(\nabla \times \bm{B}) \times \bm{B} = 0$, one arrives at the equation
\begin{equation}
    \nabla \mathcal{P} \times \nabla \mathcal{T} = 0,
\end{equation}
which implies that $\mathcal{T} = \mathcal{T}(\mathcal{P})$, and equation \eqref{eq:grad_shafranov}.
The magnetic field components can be recovered from $\mathcal{P}$ and $\mathcal{T}$ through the following relations:
\begin{align}
   B_r &= -q^2\frac{\partial \mathcal{P}}{\partial \mu} , \label{eq:B_r} \\
   B_{\theta} &= \frac{q^3}{\sqrt{1-\mu^2}}\frac{\partial \mathcal{P}}{\partial q}, \label{eq:B_theta} \\
   B_\phi &= \frac{q}{\sqrt{1-\mu^2}} \mathcal{T}\label{eq:B_phi}
\end{align}

\section{Efficient computation of the scaling parameter $\tau_k^{(1)}$} \label{app:tau_k}
The computation of inverse matrices should be avoided, since it is $\mathcal{O}(n^3)$ and also is a potential source of numerical errors if $H_k$ is ill-conditioned. 
Instead, note that from \eqref{eq:next_iterate}, \eqref{eq:direction} and \eqref{eq:sk}  we can write
\begin{equation}
    H_k^{-1} \bm{s}_k = -\alpha_k \nabla \mathcal{J}\left(\bm{\Theta}_k\right),
\end{equation}
so the explicit dependence of $\tau_k^{(1)}$ on $H_k^{-1}$ disappears. 
The step length $\alpha_k$ and the gradient $\nabla \mathcal{J} \left(\bm{\Theta}_k\right)$ are available at iteration $k$ so there is no problem in evaluating the latter expression.
Hence, the calculation of the scaling parameter only involves vector multiplications, which is $\mathcal{O}(n)$,  as
\begin{equation}
    \tau_k^{(1)} =\min \left \lbrace 1, \frac{-\bm{y}_k \cdot \bm{s}_k}{\alpha_k \bm{s}_k \cdot \nabla \mathcal{J}\left(\bm{\Theta}_k\right)} \right \rbrace.
\end{equation}

Note that the original suggestion of \cite{OrenLuenberger} for the scaling parameter was $\tau_k^{(1)} = \frac{-\bm{y}_k \cdot \bm{s}_k}{\alpha_k \bm{s}_k \cdot \nabla \mathcal{J}\left(\bm{\Theta}_k\right)}$, which was indeed motivated by the need to reduce the condition number of \eqref{eq:hessian_z_space}.
However, this choice has shown to be inferior in terms of performance compared to the standard BFGS, when combined with an inexact line search computation of $\alpha_k$, as shown in \cite{NocedalYuan} and was confirmed by our own analysis.
Instead, we followed the suggestion introduced in \cite{Al-Baali1}, which is a simple modification of the original one, but ensures theoretically super-linear convergence with inexact line searches.
It can be thought of as a switch between the standard BFGS algorithm ($\tau_k = 1$) and the self-scaled BFGS algorithm of \cite{OrenLuenberger}.
Curiously, we found that in the majority of the training iterations, the value of $\tau_k=1$ is selected, with only a small portion of them benefiting from the scaling.
Still, this was sufficient to produce considerable improvements in the optimization process.

\section{Error analysis}\label{app:errornorms}
If analytical or high-fidelity numerical solutions to the problem, denoted as $u_\mathrm{an}$, exist, the relative error of the PINN approximation $u$ is defined as:
\begin{equation}\label{eq:error_norm}
    E_u^{(2)} = \frac{\lVert u - u_\mathrm{an} \rVert_2}{\lVert u_\mathrm{an} \rVert_2},
\end{equation}
where $\lVert . \rVert_2$ indicates the $L_2$-norm evaluated in a given set of test points. 
When these are not available, we follow the method proposed in  \cite{USDP23}. To demonstrate this procedure, we use the 2D non-linear GS equation as an example. We begin by generating a uniform test grid $\left \lbrace q_i, \mu_j \right \rbrace_{i,j=1}^{N_0}$, and then evaluate the PINN solution $ \mathcal{P}_{ij} = \mathcal{P}(q_i, \mu_j, \bm{\Theta})$ at all points on the grid through a forward pass.
Then we calculate the derivatives of $\mathcal{P}_{ij}$
using a second-order finite differences scheme and construct a finite difference version of the PDE \eqref{eq:grad_shafranov}.
We define the {\it normalized} $L_2$-error of the discretized PDE as
\begin{equation}
    \epsilon_{\mathrm{FD}} 
    = \frac{1}{N_0^{d/2}} \left\lVert \triangle_{\mathrm{GS}}^{\mathrm{FD}} \mathcal{P} + \mathcal{T} \left(\frac{d\mathcal{T}}{d\mathcal{P}}\right) 
    \right\rVert_2 
    =  \frac{1}{N_0^{d/2}}\sqrt{\sum_{ij} \Big | \triangle_{\mathrm{GS}}^{\mathrm{FD}} \mathcal{P}_{ij} + \mathcal{T}_{ij} \left(\frac{d\mathcal{T}}{d\mathcal{P}}\right)_{ij}\Big |^2 }  ,
\end{equation}
where $d$ is the dimension of the problem (2D in this example).
If $\mathcal{P}_{ij}$ were the exact solution of the PDE, the error 
$\epsilon_{\mathrm{FD}}$ would decrease with increasing resolution as $\sim N_0^{-p}$, with $p$ being the order of the finite difference approximation.
In reality, $\mathcal{P}_{ij}$ is only an approximate solution with an intrinsic error $\epsilon_{\mathrm{NN}}$ depending on the (unknown) accuracy of the PINN. A reasonable way to measure $\epsilon_{\mathrm{NN}}$ is to examine the convergence of the discretized PDE residuals with resolution.
$\epsilon_{\mathrm{FD}}$ will follow the $\sim N_0^{-p}$ power law only up to the point where the PINN approximation error becomes dominant. Beyond this point,
$\epsilon_{\mathrm{FD}}$ will level off and remain roughly constant with increasing resolution, approximately equal to 
$\epsilon_{\mathrm{NN}}$. 
This behavior is clearly seen, for example, in Fig. 
\ref{fig:force_free_GS_loss_and_error}.
If the PINN approximation is inaccurate, this will happen at a relatively low resolution.
To quantify the PINN approximation error using this approach, we define 
\begin{equation}
\label{eq:eNN_definition}
\epsilon_{\mathrm{NN}} = \min \left\{\epsilon_{\mathrm{FD}} \right\}.
\end{equation}

It is worth noting that, with this strategy of using finite differences to validate the PINNs solution, the intrinsic errors due to the approximation of the derivatives are always measurable and can be reduced as much as needed (typically is a power law with increasing resolution). One can simply use a higher-order formula involving more points or even use quadruple precision if needed. This is precisely the main advantage of this validation method: it is simple and its accuracy is always under control, unlike the usual comparison with direct results from simulations which are also subject to (sometimes uncertain) errors, and there are situations where it is unclear whether the difference between the PINNs results and the numerical results are due to the PINN inaccuracy or to some other limitation in the code used as benchmark.



\bibliographystyle{elsarticle-num} 
\bibliography{bibliography}






\end{document}